\documentclass[12pt]{iopart} 
\usepackage{graphicx,color,amsmath,url,comment,xcolor}  
\usepackage[noadjust]{cite}                              
\begin{document}
\pagestyle{empty}
\title{A 1-dimensional statistical mechanics model for nucleosome positioning on genomic DNA}
\author{S. Tesoro$^1$, I. Ali$^2$, A.~N. Morozov$^3$, N. Sulaiman$^2$,  D. Marenduzzo$^3$}
\address{$^1$Theory of Condensed Matter, Cavendish Laboratory, University of Cambridge, JJ Thomson Avenue, Cambridge CB3 0HE, United Kingdom}
\address{$^2$Department of Physics, College of Science, PO Box 36, Sultan Qaboos University, Al-Khodh 123, Oman}
\address{$^3$SUPA, School of Physics and Astronomy, University of Edinburgh, Mayfield Road, Edinburgh EH9 3JZ}
\ead{st590@cam.ac.uk}

\begin{abstract}
The first level of folding of DNA in eukaryotes is provided by the so-called ``10-nm chromatin fibre'', where DNA wraps around histone proteins ($\sim$10 nm in size) to form nucleosomes, which go on to create a zig-zagging bead-on-a-string structure.
In this work we present a 1-dimensional statistical mechanics model to study nucleosome positioning within one such 10 nm fibre. {\color{black} We focus on the case of genomic sheep DNA, and we start from effective potentials valid at infinite dilution and determined from high-resolution {\it in vitro} salt dialysis experiments. We study positioning within a polynucleosome chain, and compare the results for genomic DNA to that obtained in the simplest case of homogeneous DNA,} where the problem can be mapped to a Tonks gas \cite{tonks}.
First, we consider the simple, analytically solvable, case where nucleosomes are assumed to be point-like. Then, we perform numerical simulations to gauge the effect of their finite size on the nucleosomal distribution probabilities. 
Finally we compare nucleosome distributions and simulated nuclease digestion patterns for the two cases (homogeneous and sheep DNA), thereby providing testable predictions of the effect of sequence on experimentally observable quantities in experiments on polynucleosome chromatin fibres reconstituted {\it in vitro}.
\end{abstract}


\section{Introduction}

Chromatin is the building block of chromosomes within eukaryotes~\cite{alberts,peter,understandingDNA,wolffe,schiesselreview}. It is made up by histone proteins (normally octamers) and DNA, which wraps around the histones to form a left-handed superhelix~\cite{Luger1997}. There are 147 base pairs of DNA wrapped around a histone octamer, and this complex is known as a nucleosome. 

Electron microscopy and atomic force microscopy revealed that when spread on a surface, chromatin fibres (a DNA chain containing many nucleosome) adopts a characteristic bead-on-a-string structure (Fig. 1). This is the first level of compaction of DNA within eukaryotic nuclei, which needs to be complemented by higher orders of compactions which are to date not fully understood~\cite{alberts,peter}.

\begin{figure}[ht]
 \centering
  \includegraphics[width=1\textwidth]{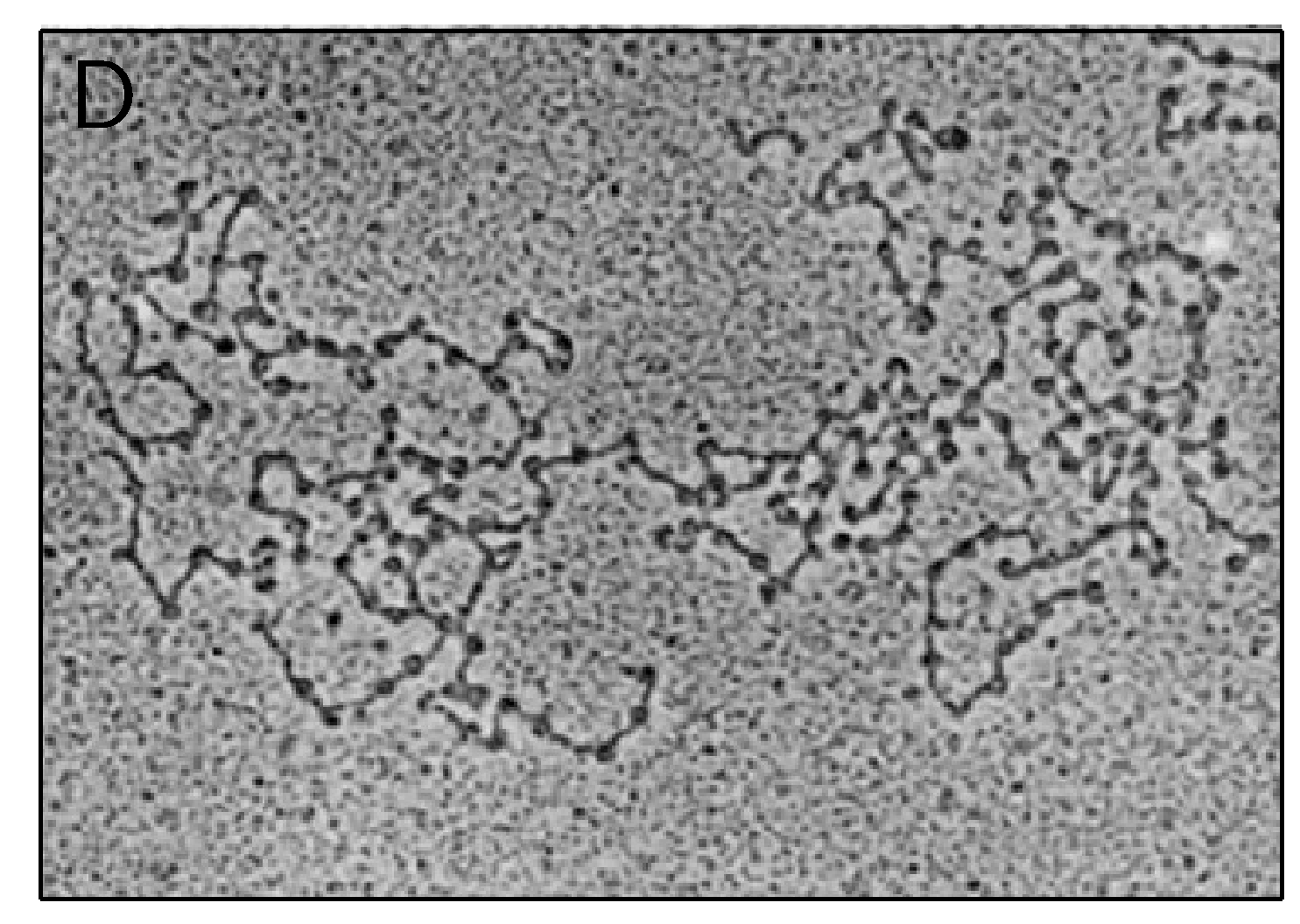}
  \caption{An electron microscopy image of a 10-nm chromatin fibre, showing the beads-on-a-string structure with well separated nucleosomes along the DNA. Reproduced with permission from Fig.~7c in Ref. \cite{Jim}.} \label{ccc}
\end{figure}

An important observation about nucleosomes is that their positioning within the genome, which has been mapped through high-throughput experiments by the genome projects, is not random, but highly reproducible. {\it In vitro}, experimentalists are also able to map the position of nucleosomes along a DNA chain of given sequence (see e.g.~\cite{invivo}). Because in the test tube there are no other constituents than DNA and histone octamers, it follows that, at least under those conditions, the positioning of the nucleosome must be dictated by simple biophysical laws: nucleosome-nucleosome interactions along the chain~\cite{jr:test}, and the nucleosome: DNA interaction. The latter interaction is partly given by electrostatic attractions between the negatively charged DNA and the positively charged histone octamer~\cite{singlenuc1,Schiessel2006,Anderson2000,schiessel}, but it also includes a sequence-specific component, which is largely due to the sequence-dependent elastic properties of the genetic material~\cite{Morozov09,inhibit,sequence,Struhl2013}.

Here we present a simple 1-dimensional model where nucleosomes diffuse along a DNA chain, interact with each other via steric repulsion, and with the genome via a sequence-dependent effective potential, which is informed by high resolution {\it in vitro} positioning experiments~\cite{jr:insilico,invivo}. As a test case, we consider a DNA sequence corresponding to part of the beta-lactoglobulin gene from the sheep genome, which was previously studied {\it in vitro} by our collaborators~\cite{invivo}. The experimental data we use {\color{black}{as an input}} considered a highly diluted situation in which, effectively, a single histone octamer was allowed to diffuse on the DNA, to find its optimal position. Nucleosomal diffusion is very slow under normal conditions~\cite{schiesselreview,Choy2012,Kulic2003}; this is why the experiments are challenging and normally require a salt dialysis protocol whereby the amount of monovalent salt in the buffer (which screens electrostatic interactions hence weakens the strong attraction between histone proteins and DNA) is slowly and very gradually decreased.

{\color{black}Our goal here is to start from these {\it mononucleosome} potentials, determined through salt dialysis experiments, and predict, either analytically or numerically, the 1D of a chromatin fibre containing a finite density of nucleosomes. Within our model, the positioning is solely due to sequence, hence our results can be seen as predictions for positioning within polynucleosome chromatin fibres of variable density which can be reconstituted {\it in vitro}. [Normally, it is more difficult to control nucleosome density {\it in vivo}, where a wealth of nucleosome positioning data exist~\cite{Kaplan09,Cheverau09,Marko07,Morozov09}.] We also simulate digestion experiments followed by gel electrophoresis, which are often used to assess quality and 1D organisation of chromatin fibre in vitro. By simulating these experiments both on the beta lactoglobulin gene and on a hypothetical homogeneous DNA where the interaction between histone octamer and DNA is uniform (sequence-independent),} we can dissect the roles of steric interactions~\cite{inhibit} and sequence~\cite{sequence,Struhl2013,Morozov09} in the positioning, at least in this specific, and highly simplified, framework. 

There are several excellent contributions available which consider the nucleosomal positioning of chromatin fibres with 1-dimensional statistical mechanics models, for instance Refs.~\cite{Kaplan09,Marko07,Cheverau09,nucleosomecit,nucleo2}.
Many of these works build on the idea that the nucleosomes along DNA behave like a Tonks gas of particles of finite size interacting via excluded volume in 1D -- this model owes its name to Lewi Tonks, who, in 1936, computed equations of state for this system \cite{tonks}.
An interesting application of Tonks' theory was provided by Ref.~\cite{jr:exact}. Such paper derives all the thermodynamical properties of DNA-protein systems in the absence of any other interaction except volume exclusion, and provides predictions for correlation functions, density functions and free energies of a 1D nucleosomal chain. Other works on chromatin building on the Tonks gas idea can be found for instance in Ref. \cite{jr:test} and Refs. therein.

A more recent work, the so-called Takashi model, is also directly relevant to our work. In the Takashi model~\cite{jr:wall}, a system of random walkers with excluded volume on a discrete lattice is considered. This model can be studied in part analytically, and some formulas for the lattice sites occupation probabilities can be derived -- however, in the generic case, these quantities need to be computed numerically. Our work can be seen as a continuation of this approach, where the sequence-spefic DNA-histone potential is included, and comes from experimental data. Under some approximations, namely when nucleosomes are point-like, but their mutual avoidance is retained, so that they cannot overtake each other on the 1D chain, and when we consider a piecewise continuous approximation of the sequence-dependent potential, we are able to solve the model analytically, and compute nucleosomal distribution functions exactly thanks to an explicit evaluation of the partition function. In the general case, we solve the model numerically with Monte-Carlo simulations.

{\color{black}The focus on analytics and exact results is one difference with respect to previous work in Refs.~\cite{Kaplan09,Marko07,nucleosomecit,nucleo2}. Another novel aspect of our work is the afore-mentioned simulation of digestion experiments~\cite{understandingDNA}}.{ \color{black}These simulations provide predictions which can be directly compared with results from micrococcal nuclease digestion of reconstituted chromatin fibres. In these experiments~\cite{understandingDNA}, chromatin fibres are subjected to the action of an enzyme (typically micrococcal nuclease) which cuts the chromatin fibre at regions of "naked" DNA (i.e., not associated with a histone octamer). The distribution of fragment length can then be measured by means of gel electrophoresis experiments on the digested fibre. Our model can, in particular, predict the effect of sequence on the output of such digestion experiments, by comparing the results obtained with genomic and homogeneous DNA.} As we shall see, these differences are quite subtle, and depend on the duration, or efficiency, of the nuclease digestion. We hope that these results may stimulate further experimental work aiming at detecting sequence-specific effects on nucleosomal positioning in reconstituted chromatin fibres.

Our work is structured as follows. In the next Section we will introduce the analytical methods used to evaluate the partition function of a 1D system of nucleosomes with a generic piecewise linear potential. We will then report, in Section 3, our analytical results on the nucleosome positional distribution functions along the chromatin fibre for the case of a uniform DNA, and for the case of the genomic DNA sequence from the beta lactoglobulin gene. In Section 4, we will go on to present our Monte Carlo simulations of a polynucleosome chain with finite size histones, focussing on the prediction for the digestion pattern. Finally, Section 5 contains our conclusions. 

\section{Methods}

\subsection{A statistical mechanics model for random walkers in a 1D potential}

In this Section we show how to extend the Tonks gas approach used in the Takashi model in Ref.~\cite{jr:wall} to the situation we are interested in. In our model, nucleosome positions are not restricted to the lattice nodes, instead, they are allowed to random walk on the continuous genomic DNA lattice.

Our aim is to calculate the statistical properties of this system. We will do so by studying random walkers with steric exclusion under the influence of an arbitrary lattice potential $V(x)$.

We begin by recalling the static approach to random walks discussed in~\cite{jr:wall} for the case in which the walkers move in a generic 1D potential on the lattice. A Boltzmann (exponential) factor in the integrand implements the effect of the potential:
\begin{equation} \label{26}
Z_N =  \int_{0}^{L}...\int_{0}^{L}e^{-\beta( V(x_1)+V(x_2)+\,...\,+V(x_N))} \theta(x_N - x_{N-1})... \theta(x_3 - x_2)\theta(x_2 - x_1)  dx_1...dx_N. 
\end{equation}

To make progress, we need to further manipulate Eq.~\ref{26} to simplify the calculations in what follows. Specifically, we observe that the $e^{-\beta (V(x_1)+V(x_2)+\,...\,+V(x_N))}$ factor in the partition function simply weighs each of the microstates a-la-Boltzmann. For simplicity, $\beta=1$ henceforth. The particular case of $V(x)=0$ for all $x$ corresponds to a uniform, sequence-independent potential, where all microstates are equally probable.

Generally, the partition function of the system is $Z_N(L)=\sum_{\{config\}}\prod_{n=1}^{N}e^{-V(x_n)} $, where configurations must include all possible sets $\{x_1,x_2,...\,,x_{N}\}$ over the lattice of length $L$, with $0 \leq x_1\leq x_2\leq ...\leq x_{N-1} \leq x_{N}\leq L$, which can be implemented introducing $\theta$-functions in Eq.~\ref{26}, ensuring sequential ordering of the particles on the lattice. The sum of ordered weighted configurations in the continuous limit gives rise to equation \ref{26}. 

By using previously defined quantities, the probability of having the $N$-th random walker at position $x_n$, which is a central quantity in our theory, can be written as follows,
\begin{equation} \label{27}
p_N(x_n=l)=e^{-V(l)}{{Z'}_{n-1}(l) {Z''}_{N-n}(L-l) \over Z_N(L)}
\end{equation}
where, ${Z'}_{n-1}(l)= \sum_{\{config\}}(\prod_{j=1}^{n-1}e^{-V(x_j)})$ and ${Z''}_{N-n}(L-l)= \sum_{\{config\}}(\prod_{j=n+1}^{N}e^{-V(x_j)})$.\footnote{Takashi had already proposed a similar algorithmic description in his work \cite{jr:wall}, but this was limited to discrete lattices.}
\newline \newline

Let's take a closer look at Eq.~\ref{27}. This equation reflects the idea that in general a partition function represents the counting of all available microstates, weighted by the Boltzmann probability. Hence, the probability of having the $n^{th}$ particle at position $l$ is the fraction of weighted microstates satisfying this constraint over all possible weighted microstates in the system.

Then, the denominator $Z_N(L)$ represents all accessible states, through the total or general partition function for $N$ particles on the lattice in Eq.~\ref{26}. The numerator of Eq.~\ref{27} represents $N$ particles on the lattice with the $n^{th}$ particle fixed at position $x_n=l$, weighted by the $e^{-V(l)}$ factor, and all weighted configurations of $n-1$ particles left of particle $x_n$ through the ${Z'}_{n-1}(l)$ expression and $N-n$ particles right of $x_n$, with the ${Z''}_{N-n}(L-l)$ expression.

The main advantage of our approach is that $Z_N(x)$,  ${Z'}_{n-1}(x)$ and  ${Z'}_{N-n}(x)$ will be analytical continuous functions in the range $\{0,L\}$. Hence, this method will allow for straightforward computations of the statistical properties of random walkers under exclusion process, such as particle PDFs or probability of a gap between particle pairs for arbitrary $N$ of particles on the lattice.

\subsection{Computing partition functions: `Divide et Impera'}

In this Section, we outline an algorithm to compute the partition functions $Z_N(L)$, which will be useful to model 1D Brownian motion under the influence of piecewise defined potentials.

One may think of a lattice of size $L$ as made up of a number $S$ of sections. Let us further consider a system of $N$ particles overall, and let us call $n_i$ the occupation number of lattice region $i$ ($1\leq i\leq S$). If we define $Z(i,n_i)$ the partition function for each lattice section $i$ and $n_i$ particles on it, the $Z(i,n_i)$ can be computed by direct application of equation \ref{26}. 

Consequently,
\begin{equation}\label{ZN2}
  Z_N(L)=\sum_{n_1=0}^{N}\sum_{n_2=0}^{N}...\sum_{n_{S}=0}^{N} \delta\left[\text{N}-\sum _{j=1}^{\text{S}} n_j\right] \prod _{i=1}^{\text{S}} Z(i,n_i)
\end{equation}
\begin{figure}[h]
  \centering
  \includegraphics[width=0.89\textwidth]{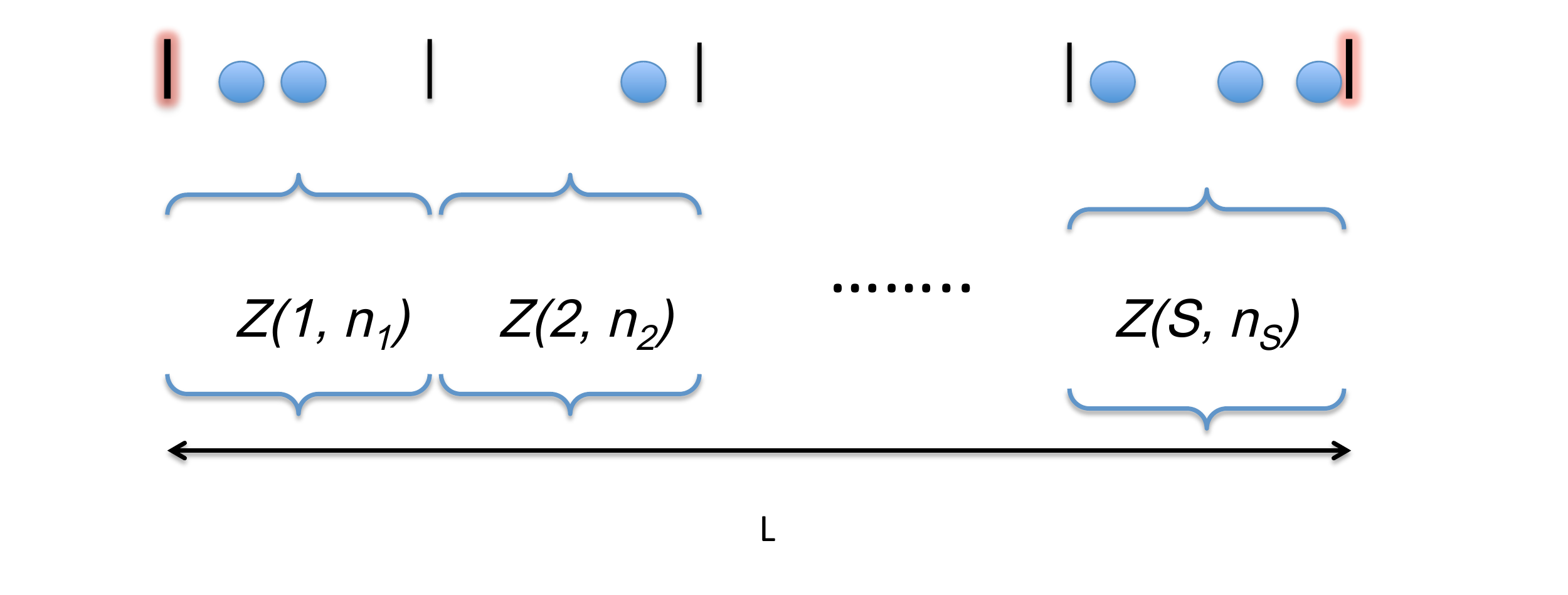}
  \caption{Product of partition functions for each lattice section. The figure represents diagrammatically one general configuration contributing to $Z_N(L)$.}\label{ztotztot}
\end{figure}

The delta function in Eq.~\ref{ZN2} ensures that there are exactly $N$ particles on the lattice. Each $Z(i,n_i)$ is a summation of weighted configurations of $n_i$ infinitesimally small particles on the $i^{th}$ lattice section. Hence, the product of all partition functions and summation over occupation numbers produce all possible microstates of the system. A diagrammatic description of equation~\ref{ZN2} is shown in Fig.~\ref{ztotztot}. 

\subsection{Indistinguishability and particle exclusion}\label{indist} 

For an arbitrary potential $V(x)$, the following observation is useful:
\begin{equation}\label{genius}
\int_{0}^{L}dx_1 \int_{0}^{L}dx_2 \,e^{-V(x_1)}e^{-V(x_2)}\theta(x_2-x_1)= {1 \over 2!}\int_{0}^{L} dx'\,e^{-V(x')} \int_{0}^{L}dx \,e^{-V(x)}
\end{equation}

The LHS counts arrangements of particles on the lattice preserving particle label ordering. This is equivalent to counting the number of possible configurations of indistinguishable particles on the lattice (RHS). Hence, a factor of $1 \over {N!}$ can be included, instead of employing the theta function. 
\if{
{\color{black} Alternatively, an argument considering the region of integration in the $x_1$-$x_2$ plane can be introduced to explain Eq.~\ref{genius}, exploiting the symmetry in the arguments of $W'(x_1,x_2)=e^{-V(x_1)V(x_2)}$. In fact, looking at Fig.~\ref{genius2}, the theta function limits the integrating region to the grey triangular area. This corresponds to reducing by ${1 \over 2!}$ the value of the integral over the whole square region of sides $L$, because of the symmetry of $W'(x_1,x_2)$.
\begin{figure}[h!]
  \centering
  \includegraphics[width=0.25\textwidth]{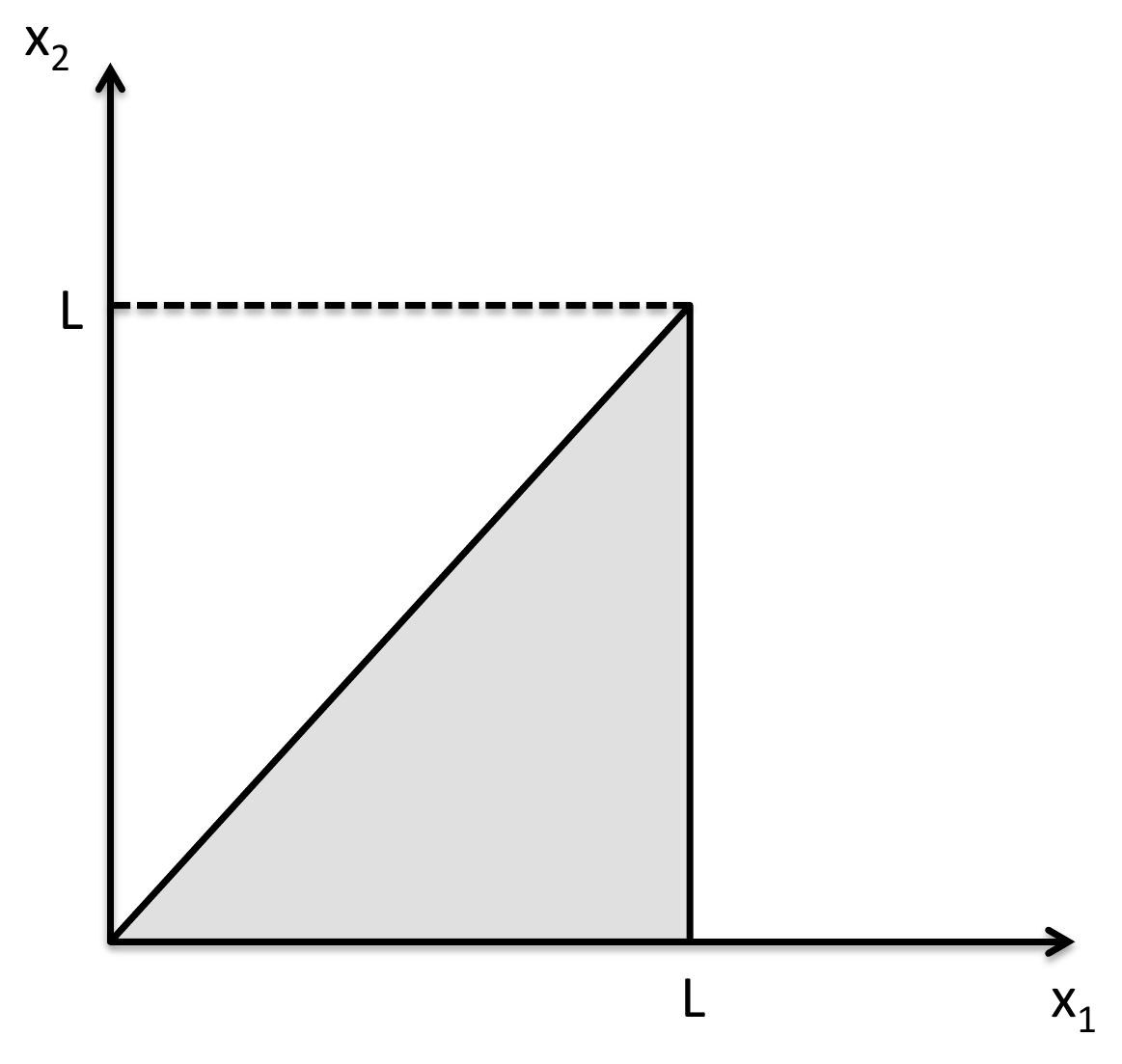}\\
\caption{\color{black}The grey region in the $x_1$-$x_2$ plane is being integrated over in the LHS of Eq.~\ref{genius}. The phase space integration area is determined by the theta function.}\label{genius2}
\end{figure}}
}\fi

Generalising these ideas to $N$ particles, Eq.~\ref{26} can be recast as
\begin{equation}\label{simp}
Z_N = {1 \over N!}{ \left( \int_{0}^{L}dx\, e^{-V(x)}\right)}^N ={1 \over N!}(Z_1)^N.
\end{equation}
Consequently, equation \ref{27} now becomes
\begin{equation}\label{boson}
p_N(x_n=l)=e^{-V(l)}{({Z'}_{1}(l))^{n-1}({Z''}_{1}(L-l))^{N-n} \over {Z_N(L)(n-1)!(N-n)!}}.
\end{equation}
Here we explain how to compute  $Z'$ and $Z''$ for piecewise potentials exploiting the algorithmic approach in Eq.~\ref{ZN2} and implementing the result in Eq.~\ref{simp}.

In the following, a special class of potentials will be studied: piecewise linear potentials, so that there are constant gradients in the potential for each lattice section.
{Figure \ref{example} shows a graphical representation of a possible configuration of the system and its relative representation in terms of partition functions for the various lattice sections.

\begin{figure}[h!]
  \centering
  \includegraphics[width=0.89\textwidth]{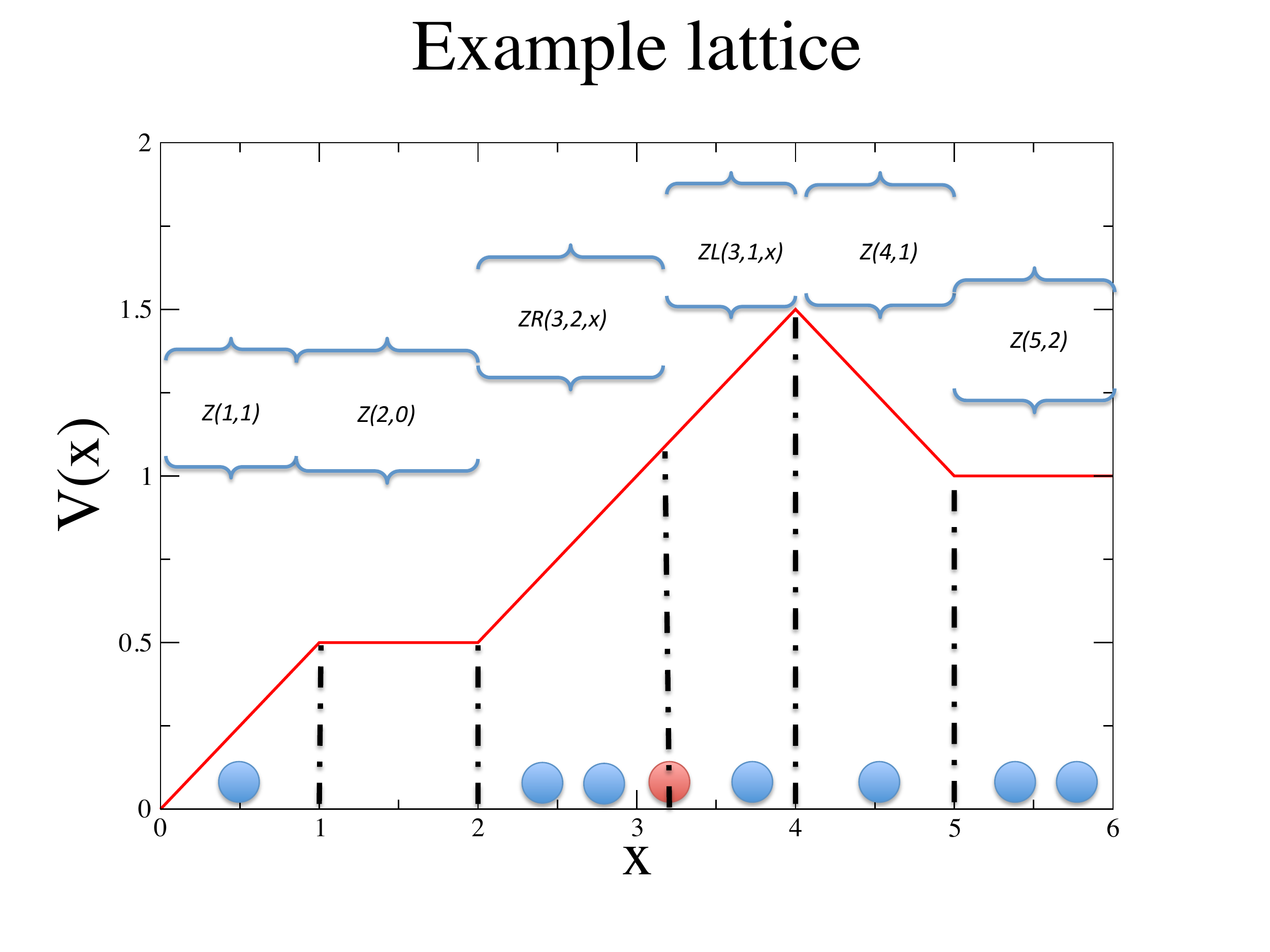}
  \caption{ Computation of Eq.~\ref{27} exploting equation \ref{ZN2} in order to compute the probability distribution function for the red particle at position $X$.} \label{example}
\end{figure}

In this case, the partition function for section $i$ will be determined by  $L_i$, the length of the lattice section, and $m_i$, the constant gradient of the potential over the region with $n$ particles ($V(0)$ is the value of the potential at the origin of the lattice section), as follows:
 \begin{small}
 \begin{equation}\label{simp3}
Z(i,n)={1 \over N!}(Z(i,1))^n ={1 \over N!}\left( \int_{0}^{L_i} e^{-m_ix+ V(0)}~ dx\right)^n= {1 \over N!}\left({e^{ V(0)}\over m_i} -{e^{-m_iL_i + V(0)}\over m_i} \right)^n.
 \end{equation}
 \end{small}
Finally, the probability for particle $l$ to be at position $x$ within section $i$ can be written as:
  \begin{small}
    \begin{equation}\label{algorithm}
           p_N(x_l=x)= {e^{-V(i,x)} \over { Z_N{N-l}!}{l-1}!}  {  \bigg(\sum_{j=1}^{i-1} Z(j,1) + ZL(i,1) \bigg) }^{l-1} { \bigg( \sum_{k=1+i}^{S} Z(k,1) + ZR(i,1) \bigg) }^{N-l}   
\end{equation}
  \end{small}
  where $ZR(i,n)$ and $ZL(i,n)$ are the partition functions representing configurations with $n$ particle to the right or left of particle $l$ within section $i$ of the lattice. Such functions are defined through equation \ref{simp3}.

  Eq.~\ref{algorithm} is obtained {\color{black} by substituting Eq.~\ref{simp3} into Eq.~\ref{27} where appropriate. Computations via this recast version of Eq.~\ref{27} are faster in practice. This is because Eq.~\ref{algorithm} now involves powers of single particle partition functions,} which can be determined algorithmically through the formula on the RHS of Eq.~\ref{simp3} for this particular class of piecewise potentials.
}

These algorithms are the main novelty of our analytical approach, and they dramatically reduce the computational time needed to get an explicit expression for nucleosomes position probability distribution functions (PDFs). {\color{black} For instance, implementing these algorithms through 'Wolfram Mathematica', it takes less than 5 seconds for our code to output PDFs for 30 particles on the genomic DNA lattice}. This is the method we use to compute the results presented in the next Section.

\section{A case study: point-like nucleosomes on a homogeneus and on a genomic DNA}

\begin{figure}[h!]
  \centering
  \includegraphics[width=0.89\textwidth]{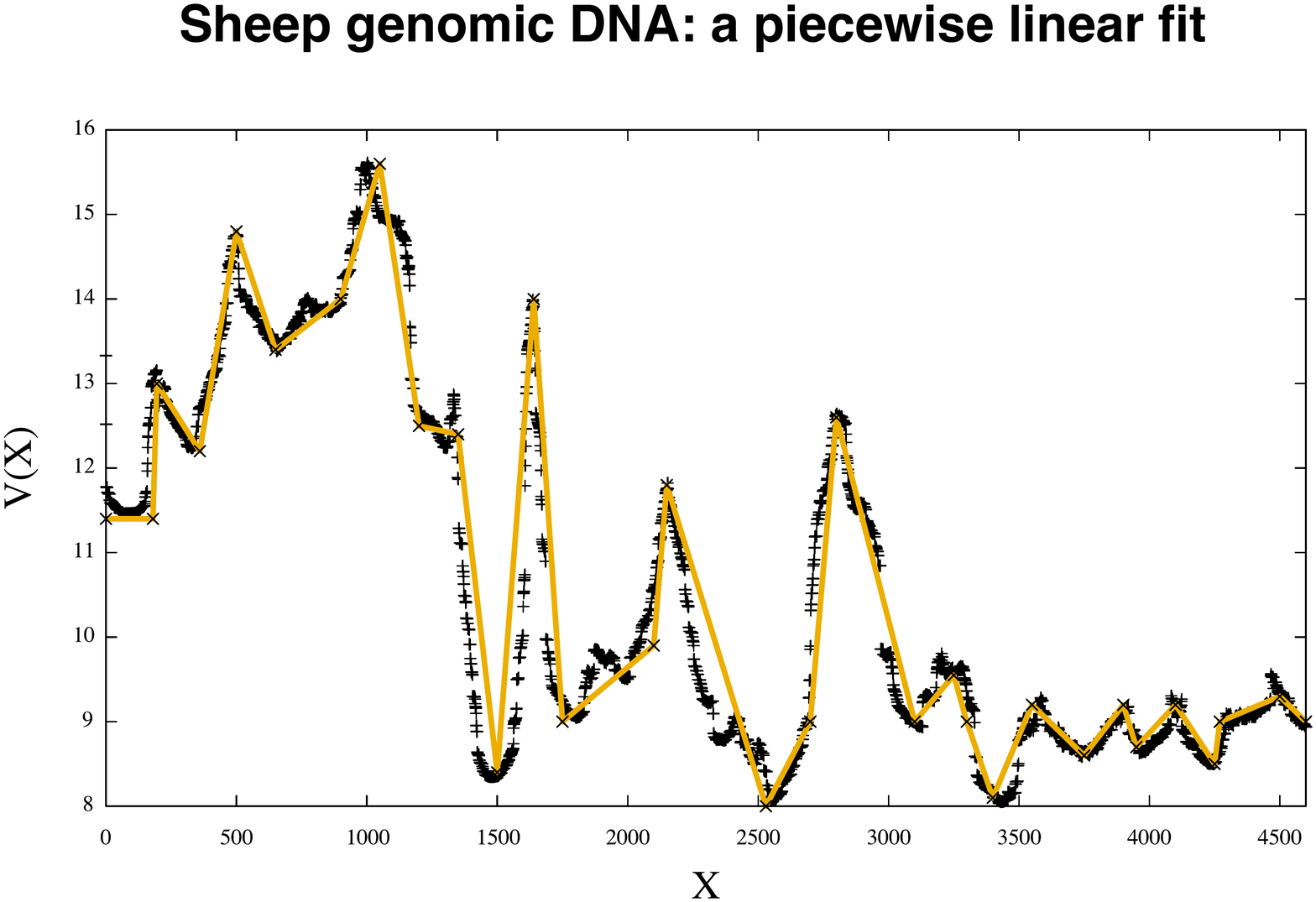} 
  \caption{This plot shows the DNA potential as inferred from (mono)nucleosome positioning experiments on a genomic segment from the {\color{black} beta lactoglobuline gene from sheep DNA. In yellow a piecewise linear fit to the genomic DNA potential}.} \label{DNApotential111}
\end{figure}

In this Section we provide a series of analytical results for nucleosome positioning within: (i) a genomic DNA segment (the beta lactoglobuline gene in the sheep); and (ii) a homogeneous DNA as a reference case. As we shall see, we are here able to obtain explicit results for an arbitrary (sequence-dependent) histone:DNA interaction potential which is piecewise linear -- equivalently, a potential with piecewise constant first derivative, or gradient. This approximation is useful as it naturally breaks up the DNA molecule into $S$ sections, within each of which the potential is linear. As can be seen in Fig.~\ref{DNApotential111}, the case of the genomic DNA segment (which comes from the observed positions of nucleosome dyads, i.e. centres) can be approximated fairly well with a piecewise linear function. 

\subsection{Positional probability distribution function for a chain of ten nucleosomes}

In this Section we plot our analytical solution for the positional probability distribution function (PDFs) of each of the nucleosome in a chain where 10 histone octamers deposit on the first 4.6 kilo-base pairs of the sheep beta lactoglobuline gene (explicit expressions were derived using a `Wolfram Mathematica' code based on Section \ref{indist}). The situation we consider corresponds to a coverage of one nucleosome per 460 base pairs, which is about half the physiological one, where one nucleosome corresponds to about 200 base pairs~\cite{alberts,understandingDNA}. This low density is chosen so as to avoid crowding effects which are likely to invalidate our approximation of point-like nucleosomes which we use in our analytics.

\begin{figure}[h!]
  \centering
  \includegraphics[width=.45\textwidth]{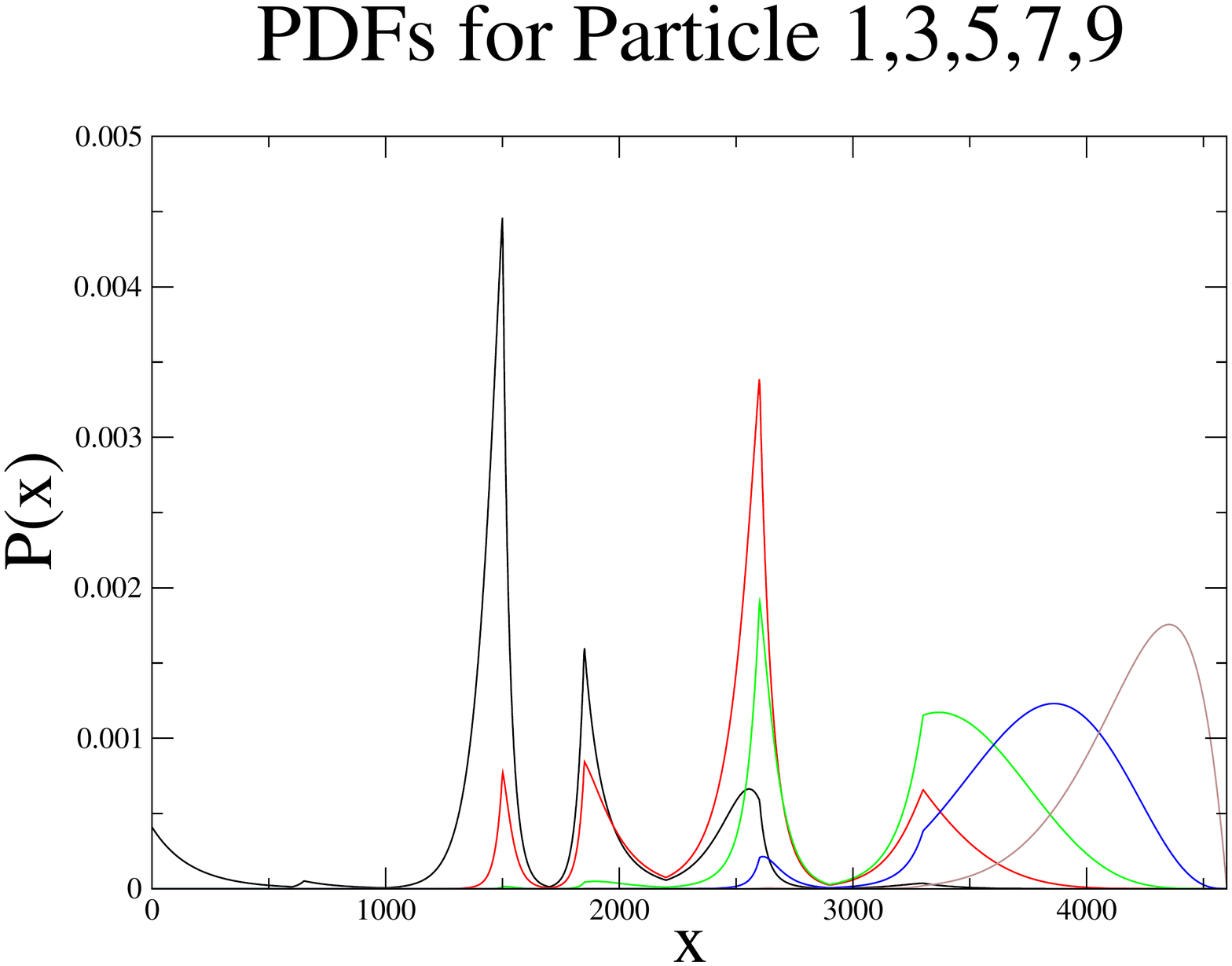}
    \includegraphics[width=.45\textwidth]{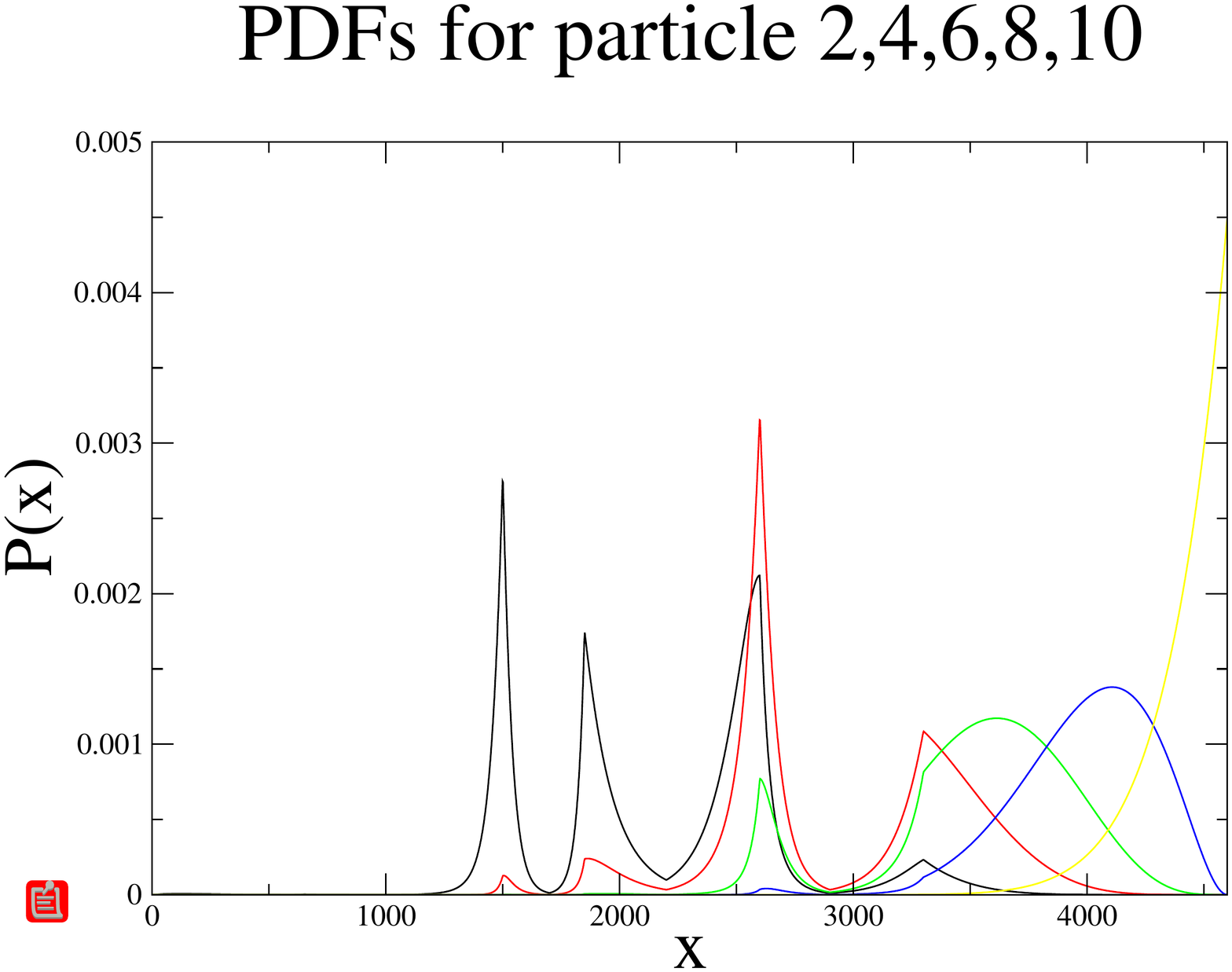}
    \caption{Plots of the nucleosomal positional PDFs for 10 particles on 4600 DNA bases (part of the beta lactoglobulin gene). {\color{black} Colours correspond to: particle 1 and 2 (black), particle 3 and 4 (red), particle 5 and 6 (green), particle 7 and 8 (blue), particle 9 (brown), particle 10 (yellow)}. }


\label{PDFsheep1}
\end{figure}
\begin{figure}[h!]
  \centering
  \includegraphics[width=.45\textwidth]{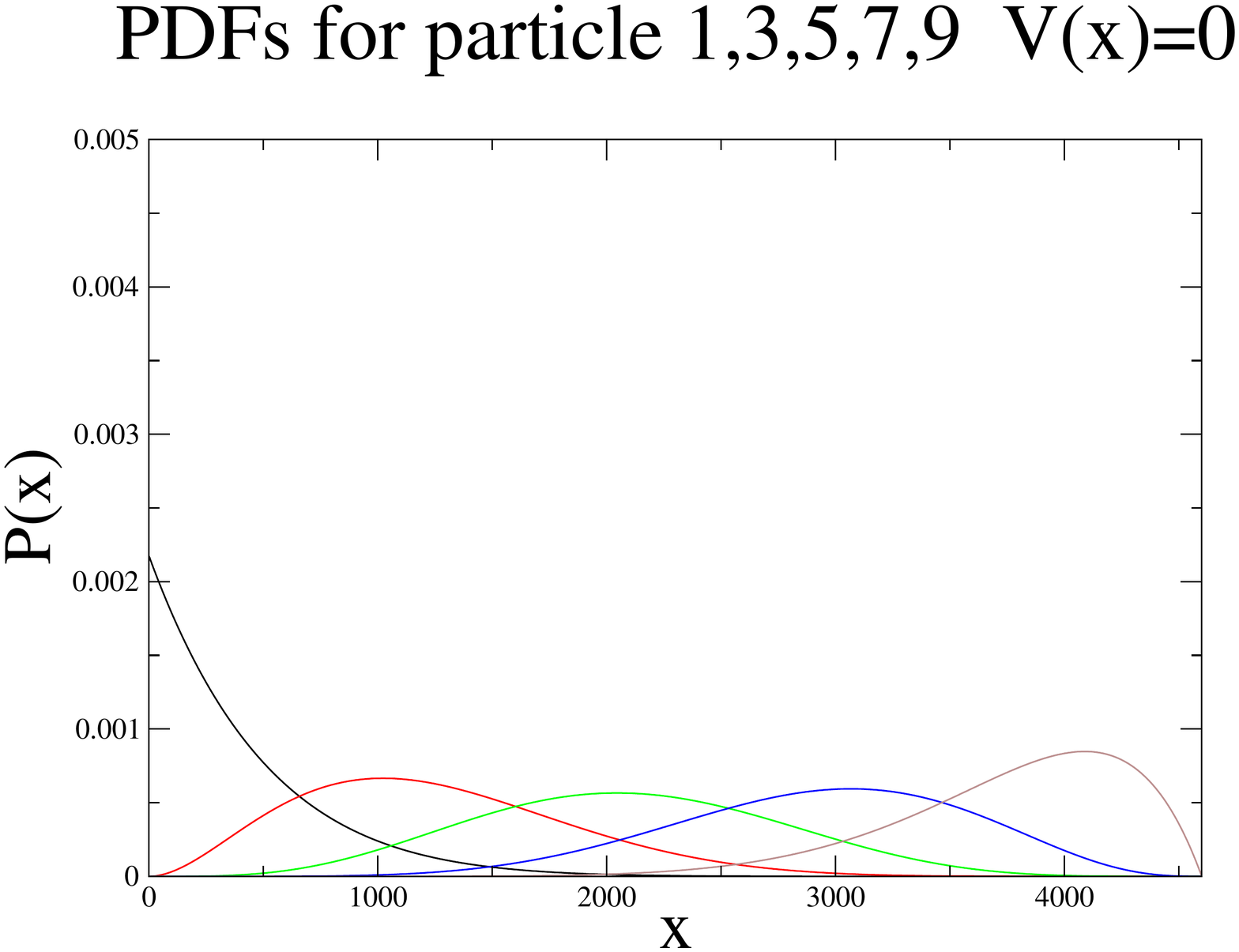}
  \includegraphics[width=.45\textwidth]{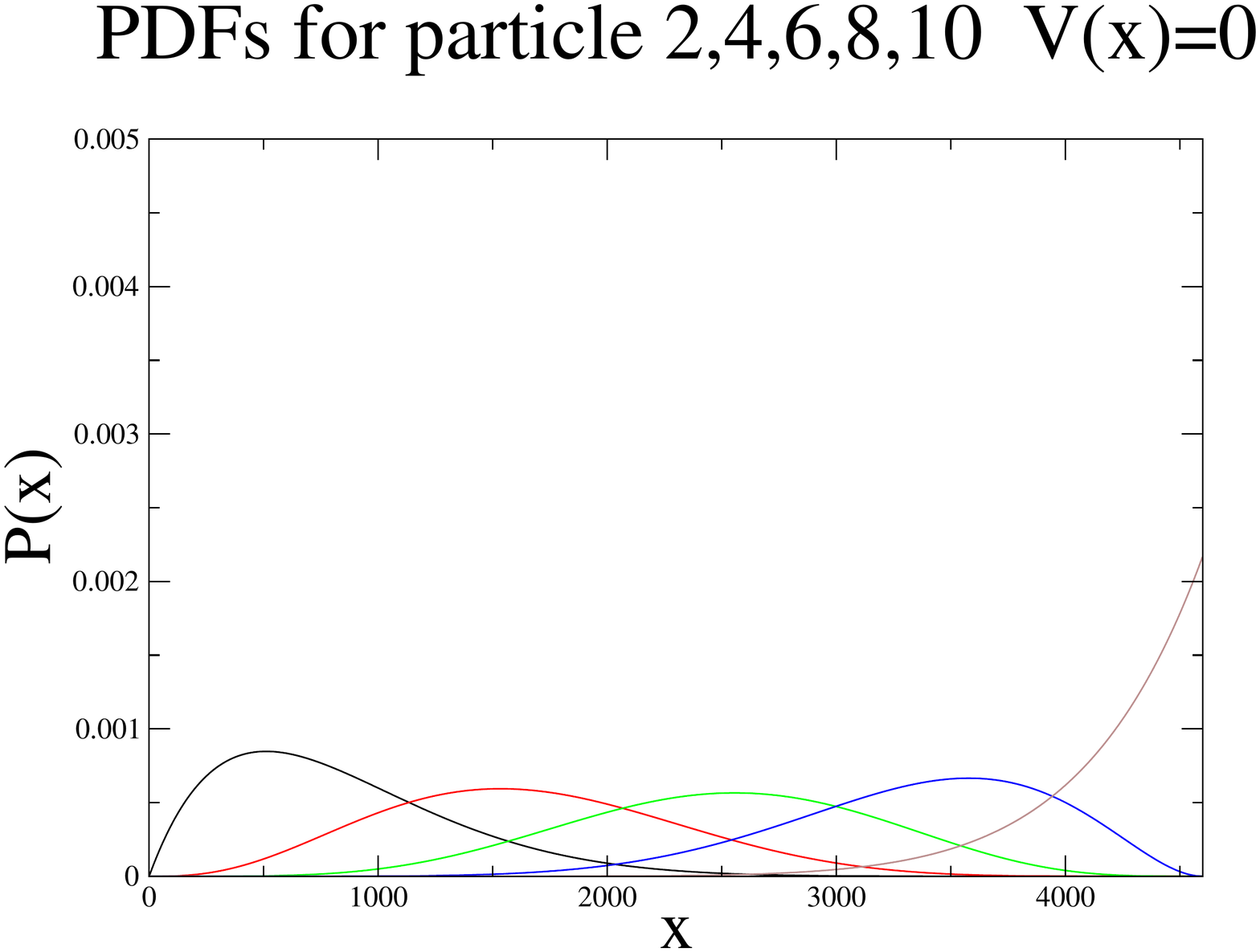}     
  \caption{Plots of the nucleosomal positional PDFs for 10 particles on a homogeneous DNA. Colours correspond to: {\color{black} particle 1 and 2 (black), particle 3 and 4 (red), particle 5 and 6 (green), particle 7 and 8 (blue), particle 9 and 10 (brown)}.}  
\end{figure}
\begin{figure}[h!]
 
\label{PDFsheep0}
\end{figure}

Fig ~\ref{PDFsheep1} show the position of the odd and even nucleosomes/particles. Some observations can be made about the most likely positions of the nucleosomes predicted analytically. First, the histone:DNA experimentally obtained potential is higher for the first 1000 bases than for the rest of the molecule. Hence, except for particle 1, no other particle presents a significant probability of localising within such a region. Second, again in line with intuition, we observe that the particles' PDFs have their highest peaks in lattice regions corresponding to troughs or deep minima in the potential. Third, it is interesting to note that, for neighbouring particles, the maxima and minima of the PDFs tend to coincide, but they vary in heights. For instance, the PDFs for particle 1 and particle 2 both have peaks at $x\sim 1500$ and $x\sim 1900$ (base pairs), but at $x\sim 1500$ the peak for particle 1 is higher than for particle 2, while at $x=1900$ the situation in reversed. This is caused by the steric, or exclusion, interaction between the nucleosomes, as their order is fixed within the chromatin fibre (they cannot overtake each other). To highlight the effect of sequence, we compare in Fig.~\ref{PDFsheep0} the PDFs for nucleosomes in the case of the genomic DNA segment with the case of a homogeneous DNA where there is no sequence-dependent variation in the histone:DNA interaction. It can be seen that, at least at this coverage, sequence makes a large difference in determining the nucleosome positional PDFs within a chromatin fiber.
  
\subsection{Gap probabilities for 10 particles on the DNA lattice}\label{results}

Another interesting quantity we can obtain from our theory are the `gap probabilities', which is the probability of a `gap' of $g$ base pairs between successive nucleosomes. This quantity is, as we will discuss more in detail later on, relevant for digestion experiments with nuclease. In the context of this analytical calculations, such gap PDFs are useful to evaluate the limitations in our approach, as gaps in reality need to be larger than the nucleosome size, i.e. 146 base pairs, which is disregarded in our analytics (which assumes point-like nucleosomes, {\color{black} the interested reader will find more results for gap probabilities in a Tonks gas at \cite{reftonks1} and \cite{reftonks2}}). 

\begin{figure}[h!]
  \centering
  \includegraphics[width=0.7\textwidth]{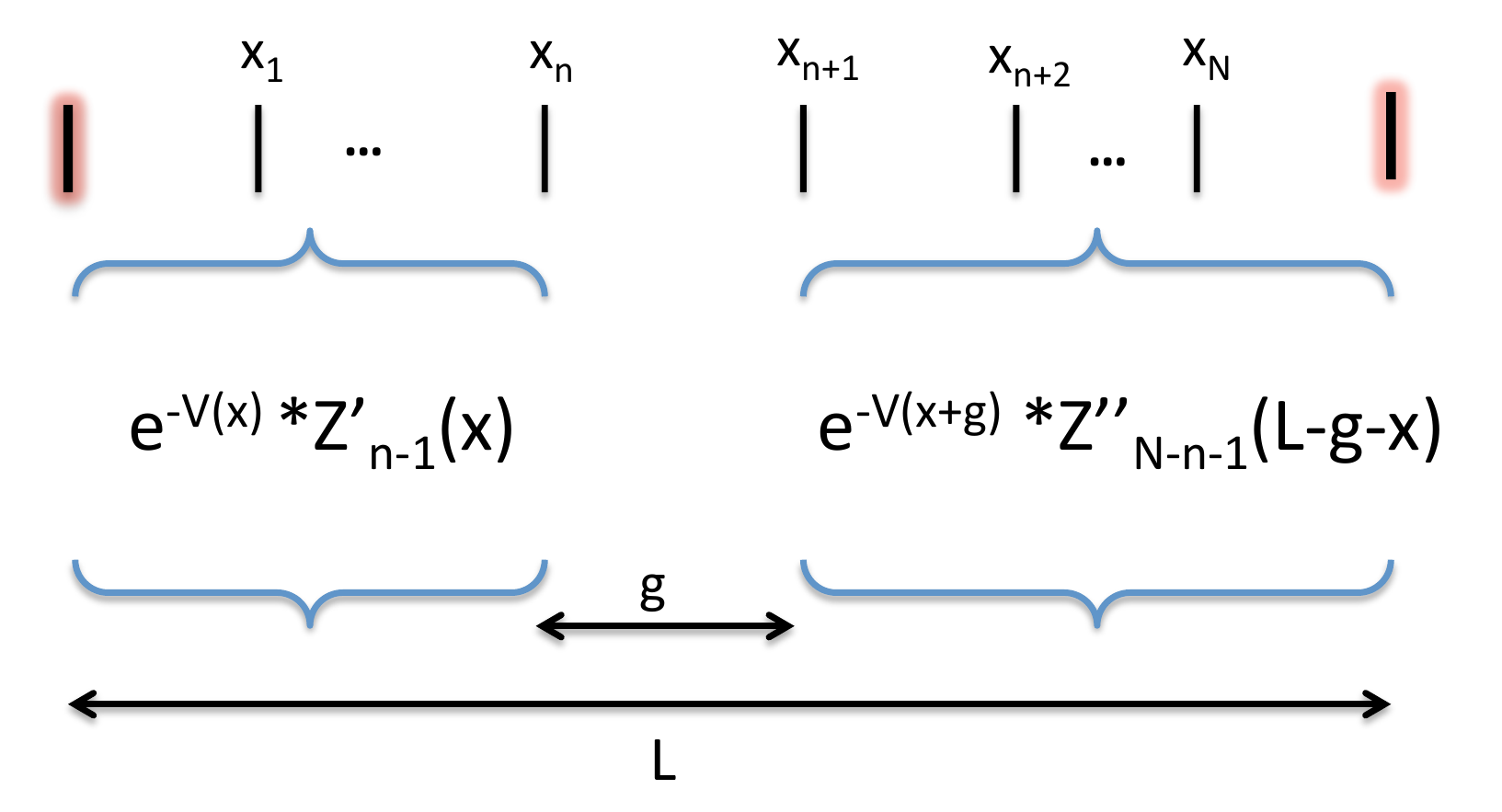}\\
\caption{Graphical description of the terms in Eq.~\ref{inculata}.}\label{gap69}.
\end{figure}

How can we compute analytically the probability of a gap of size `$g$' between particle `$n$' and particle `$n+1$', with particle `$n$' at position `$x$' and particle `$n+1$' at position `$x+g$'? The route we follow is to perform a normalised sum of all microstates which satisfy the condition of having a gap of $g$ base pairs between the $n$-th and the $(n+1)$-th particles. The relevant formula for this gap PDF is therefore
\begin{equation}\label{inculata}
P_{\rm gap}(n,n+1,g,x)={{Z'}_{n-1}(x)e^{-V(x)}e^{-V(x+g) }{Z''}_{N-n-1}(L-x-g)\over Z_N},
\end{equation}
where ${Z'}_{n-1}(x)$ represents weighted configurations of particles located to the left of particle `$n$' and ${Z''}_{N-n-1}(L-x-g)$ represents weighted configurations of particles located to the right of particle `$n+1$'. The $e^{-V(x)}$ and $e^{-V(x+g)}$ factors take into account the configurational weights of particle `$n$' and particle `$n+1$' on the lattice. Figure \ref{gap69} describes the role of the functions in Eq.~\ref{inculata} \footnote{$(Z'')_{N}(x)$ and $(Z')_{N}(x)$ can be computed either through methods previously shown or much more simply through algorithms presented in section \ref{indist}.}.

\begin{figure}[h!]
    \includegraphics[width=0.45\textwidth]{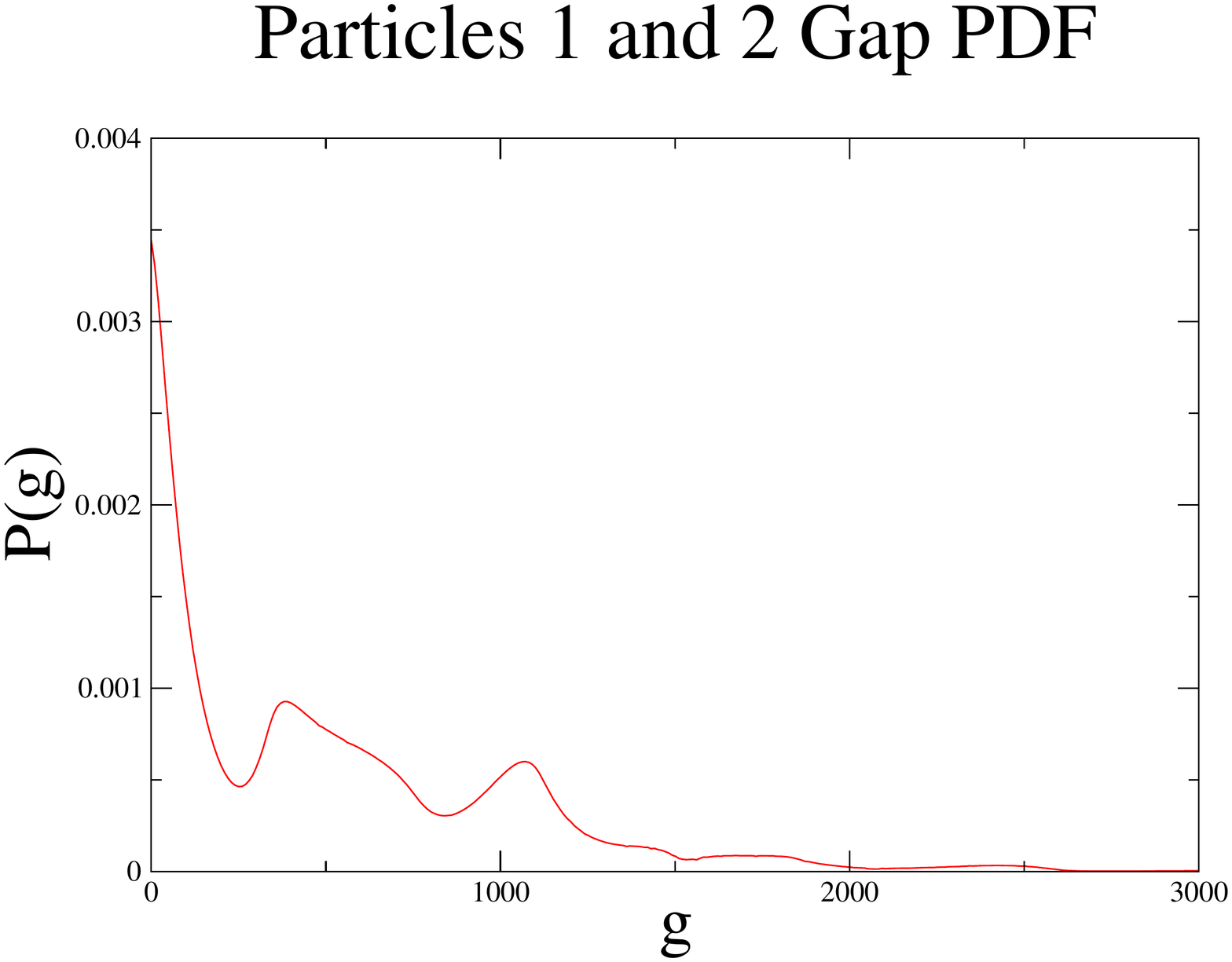}
    \includegraphics[width=0.45\textwidth]{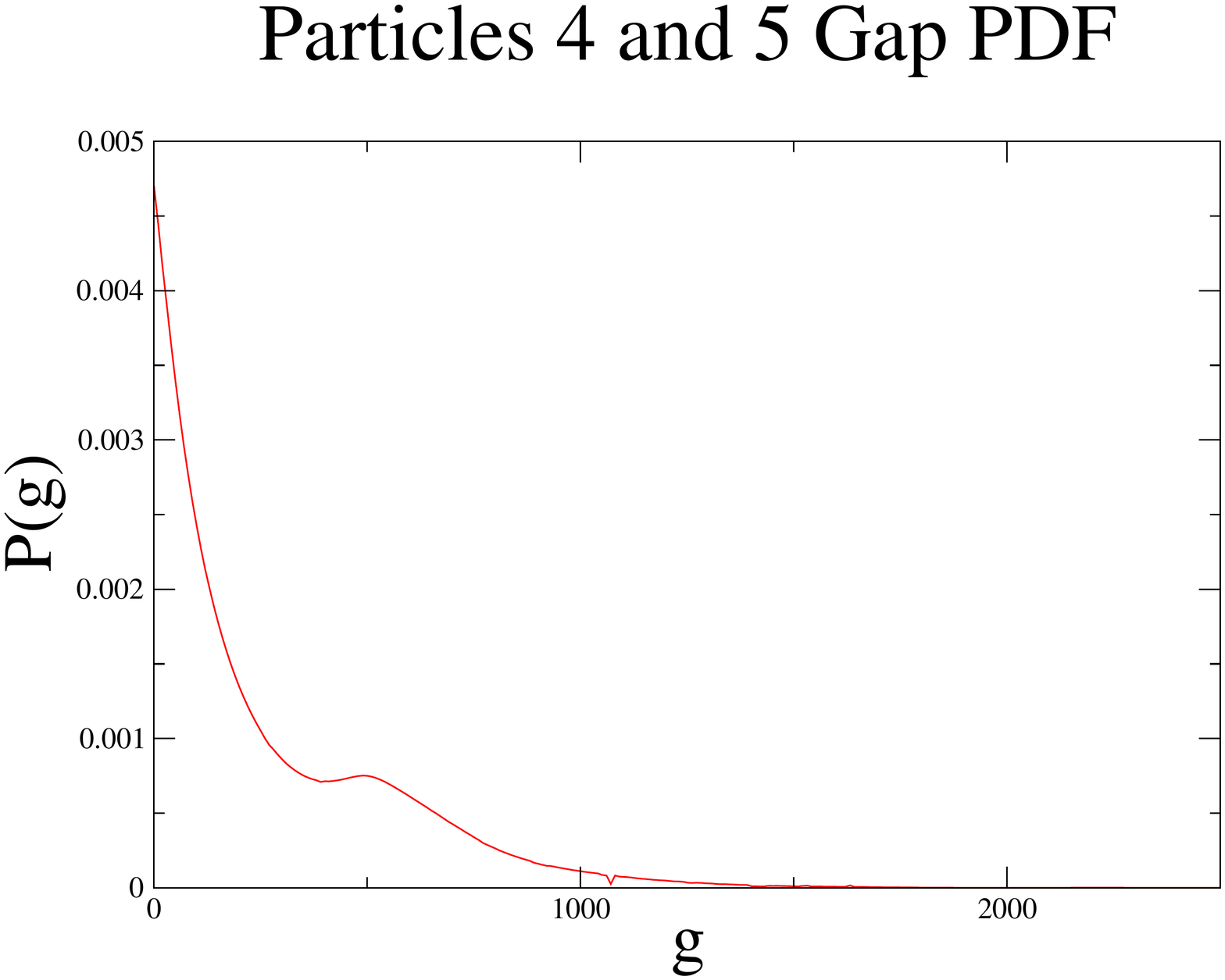}
  \centering
    \includegraphics[width=0.45\textwidth]{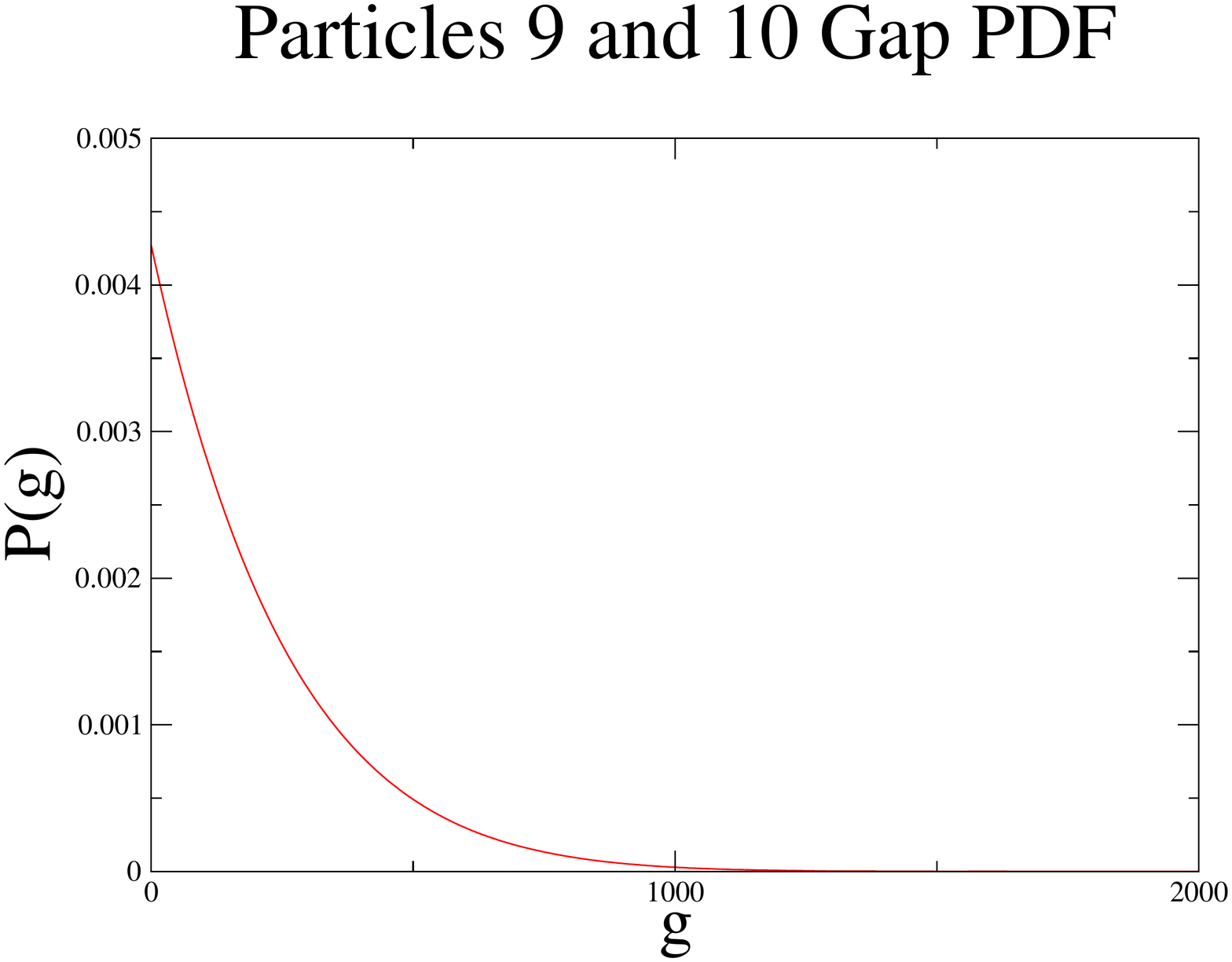}\\
  \caption{Gap probabilities for particle 1 and 2, particle 4 and 5, particle 9 and 10. For the plots shown, there is a total of 10 particles on the DNA lattice.}\label{GAPP}
\end{figure}

The following equation,
\begin{equation}\label{integrate}
PDF_{\rm gap}(n,n+1,g)=\int_0^{L-g} P_{\rm gap}(n,n+1,g,x)\, dx
\end{equation}
instead describes the probability for particle `$n$' and particle `$n+1$' to display a gap of size `$g$' {\it anywhere} on the lattice ($L$ is the total number of base pairs). Through numerical integration, gap PDFs have been produced and are shown in Fig.~\ref{GAPP}. These have been evaluated and plotted extending the  `Wolfram Mathematica' code introduced in the previous section. 

From figure \ref{GAPP}, one can readily notice that there is a finite probability for neighbouring particles pairs to have a gap of less than 147 DNA base pairs. Therefore, it is clear that our approximation will be inaccurate for dense nucleosomes systems; however it can still provide a good approximation for dilute systems (see also below, where we compare with numerical simulations incorporating realistic nucleosome sizes). {\color{black} If particles had finite sizes, then $P(0)=0$, however average gap sizes $<g>$ are expected to be line with more realistic scenarios.}


\section{The effect of nucleosome size: Monte-Carlo simulations of nucleosomes on homogeneous and genomic DNA}

In the previous Section, we studied the statistics of point-like nucleosomes on a polynucleosome chromatin fibre, taking into account steric nucleosome-nucleosome interactions by a simple exclusion interactions (nucleosomes could not overtake each other along the DNA chain). It is obviously of interest, especially when the density of nucleosome along the fibre is large, to relax this approximation. In this Section we will therefore present Monte-Carlo simulations of the positioning of nucleosomes along a DNA, again considering the cases of a homogeneous DNA and of the genomic DNA segment, as done in Section 3 analytically. Before we present the results of these simulations, we briefly discuss, as an aside, how one could generalise our previous exact treatment to correctly consider the finite nucleosome size. As this avenue requires ultimately a numerical treatment, we focus later on direct Monte-Carlo simulations. 
\newline {\color{black} We note that limitations to an equilibrium approach to Monte Carlo simulations of nucleosome position patterns (especially when used to study chromatin {\it in vivo}) are discussed in~\cite{nucleosomecit}. We also note that Monte-Carlo simulations similar to those reported here can be found in \cite{nucleosomecit,Marko07,nucleo2}.}

\subsection{A possible exact treatment for excluded volume between nucleosomes on the DNA lattice}

Consider having a single nucleosome on a DNA of length L. Then, the weight of a general microstate with the particle occupying the lattice region between $x$ and $x+c$ ($c$ is the size of the particle) is:
\begin{equation}
W(x)=\exp\left({-\int_{x}^{x+c}V(x)\,dx}\right)
\end{equation}
To show this, first evaluate $W(x)$ for a discrete lattice and then let $\Delta x \rightarrow 0$ ($\Delta x$ is the lattice spacing) to represent a continuous lattice:  
\begin{equation}\label{Volterra}
W(x)= \lim_{\Delta x \to 0} \prod_{n=x}^{x+c} e^{-V(n)\Delta x} = \lim_{\Delta x \to 0} \prod_{n=x}^{x+c} e^{ \ln (e^{-V(n)\Delta x})} = \exp\left({\int_{x}^{x+c}\ln\left(e^{-V(x)}\right)\,dx}\right) 
\end{equation} 
Then, the partition function becomes the sum of all weighted microstates.
\begin{equation} 
Z_1=\int_{0}^{L-c}W(x)\,dx=\int_{0}^{L-c}\exp\left({-\int_{x}^{x+c}V(x')\,dx'}\right)\,dx
\end{equation}

The generalisation for $Z_N$ is as follows:
\begin{small}
\begin{equation}\nonumber
 \begin{gathered}\label{hardhard}
   \begin{split}
     Z_N=\int_{0}^{L}dx_1\int_{0}^{L}\,dx_2\,...\, \int_{0}^{L}\,dx_{N-1}\int_{0}^{L-c}\,dx_N W(x_1)W(x_2)\,...\,W(x_N)\\ \theta(x_N - x_{N-1}-c)\,...\, \theta(x_2 - x_{1} -c)\\
     =\int_{0}^{L}\,dx_1 \,...\, \int_{0}^{L}\,dx_{N-1}\int_{0}^{L-c}\,dx_N\,\exp\left({-\int_{x_1}^{x_1+c}V({x}_1^{'})\,{dx}_1^{'}}\right)\,...\,\\ \exp\left({-\int_{x_N}^{x_N + c}V({x}_N^{'})\,{dx}_N^{'}}\right)\theta(x_N - x_{N-1}-c)\,...\,\theta(x_2 - x_{1}-c)
   \end{split}
 \end{gathered}
\end{equation}
\end{small}
\newline
While these integrals provide an explicit route to the evaluation of the partition function, and ultimately of the positional PDFs for nucleosomes, they cannot easily be evaluated for a large number of particles, even through numerical methods. Therefore, in what follows we present results obtained from Monte-Carlo simulations on a 1D lattice where the finite size is directly considered, together with the effect of the sequence-specific DNA:histone interaction. 

\subsection{Monte-Carlo algorithm, and its validation}

Our Monte-Carlo approach consists of dynamical simulations where $N$ particles of finite size (146 base pairs) diffuse on a lattice of $L$ base pairs ($L=4600$ when comparing to the analytics, $L=10943$ otherwise -- the latter value corresponds to the whole beta lactoglobulin gene). We consider variable nucleosome density, corresponding to $10\le N \le 30$. Each nucleosome interacts with the DNA through volume exclusion (the histones are either pointlike, or with size equal to 147 base pairs) and one of the following: (i) either a sequence-dependent potential obtained from the experiment in Ref.~\cite{invivo} for the genomic DNA sequence extracted from the beta lactoglobulin gene, (ii) or a sequence-independent potential corresponding to a homogeneous DNA. As in Section 3, the sequence-dependent potential, $V(x)$, is derived from the single-nucleosome positional PDFs $p(x)$, mapped experimentally in Ref.~\cite{invivo}, as $V(x)=-k_BT \log\left({p(x)}\right)$.
For the boundary conditions, when wanting to compare with our analytics ($N=10$ and $L=4600$ base pairs, Section 4.2 and 4.3), we consider an open DNA chain with reflecting boundary conditions at the ends; in particular, nucleosomes are not allowed to move if they are at one of the end of the chains and the trial move will cause them to hop off the chain. Instead, in Section 4.4 we use periodic boundary conditions, corresponding to a DNA loop (as used in chromatin reconstitution experiments {\it in vitro}~\cite{invivo}).

Initially, all $N$ nucleosomes are dispersed on the DNA lattice uniformly, at a fixed mutual distance (results did not depend on the initial condition provided the simulation was long enough). For each time step, we then selected randomly one of the nucleosomes, and moved it to the right or left (with equal probability). Provided that the move does not lead to a steric clash with neighbouring nuclesomes, we then accepted or rejected the move according to the Metropolis criterion, which depends on the potential which is chosen.{ \color{black}Similar algorithms are used in \cite{nucleosomecit,Marko07,nucleo2}.}

\begin{figure}[h!]
  \centering
  \includegraphics[width=0.45\textwidth]{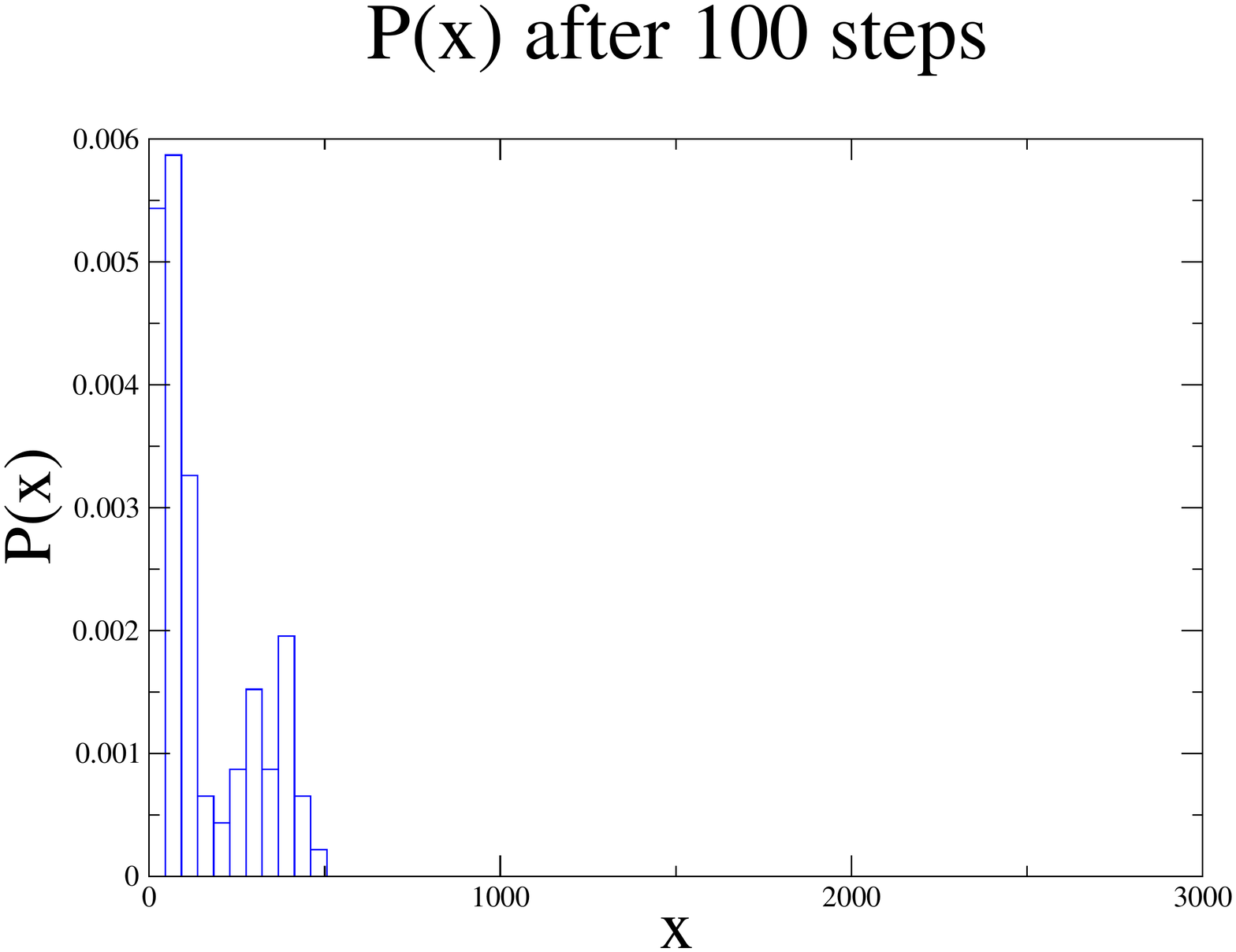}
  \includegraphics[width=0.45\textwidth]{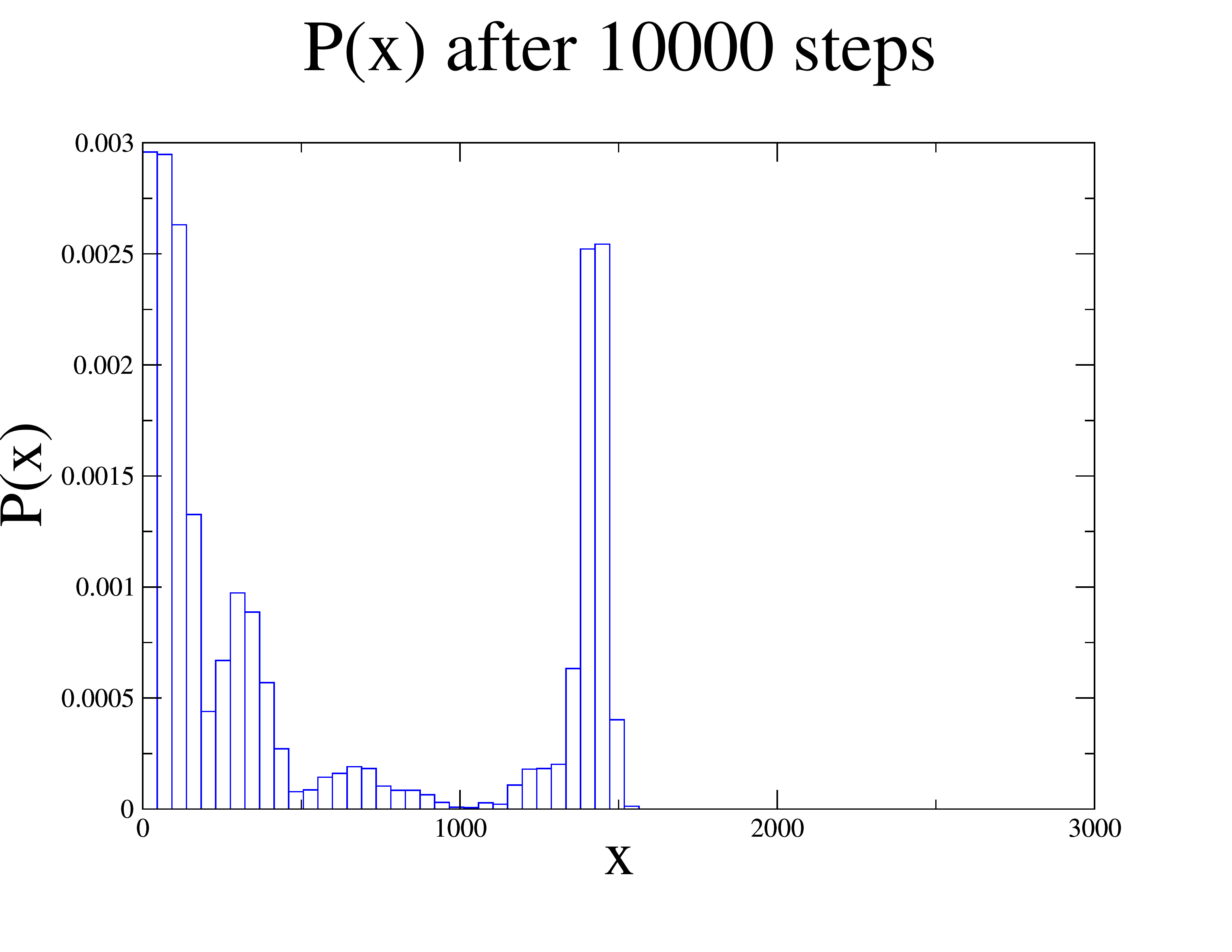}\\
  \includegraphics[width=0.45\textwidth]{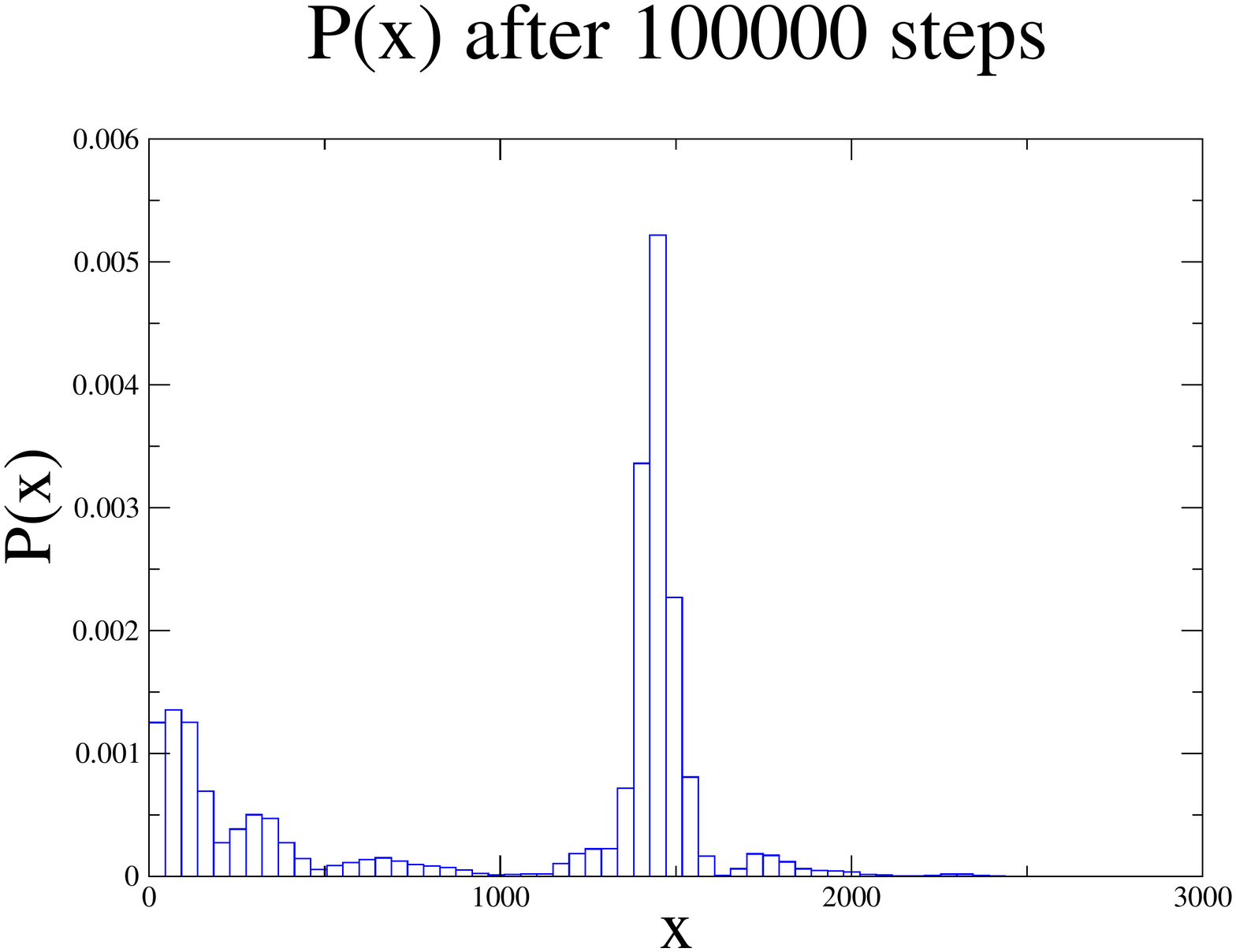}
  \includegraphics[width=0.45\textwidth]{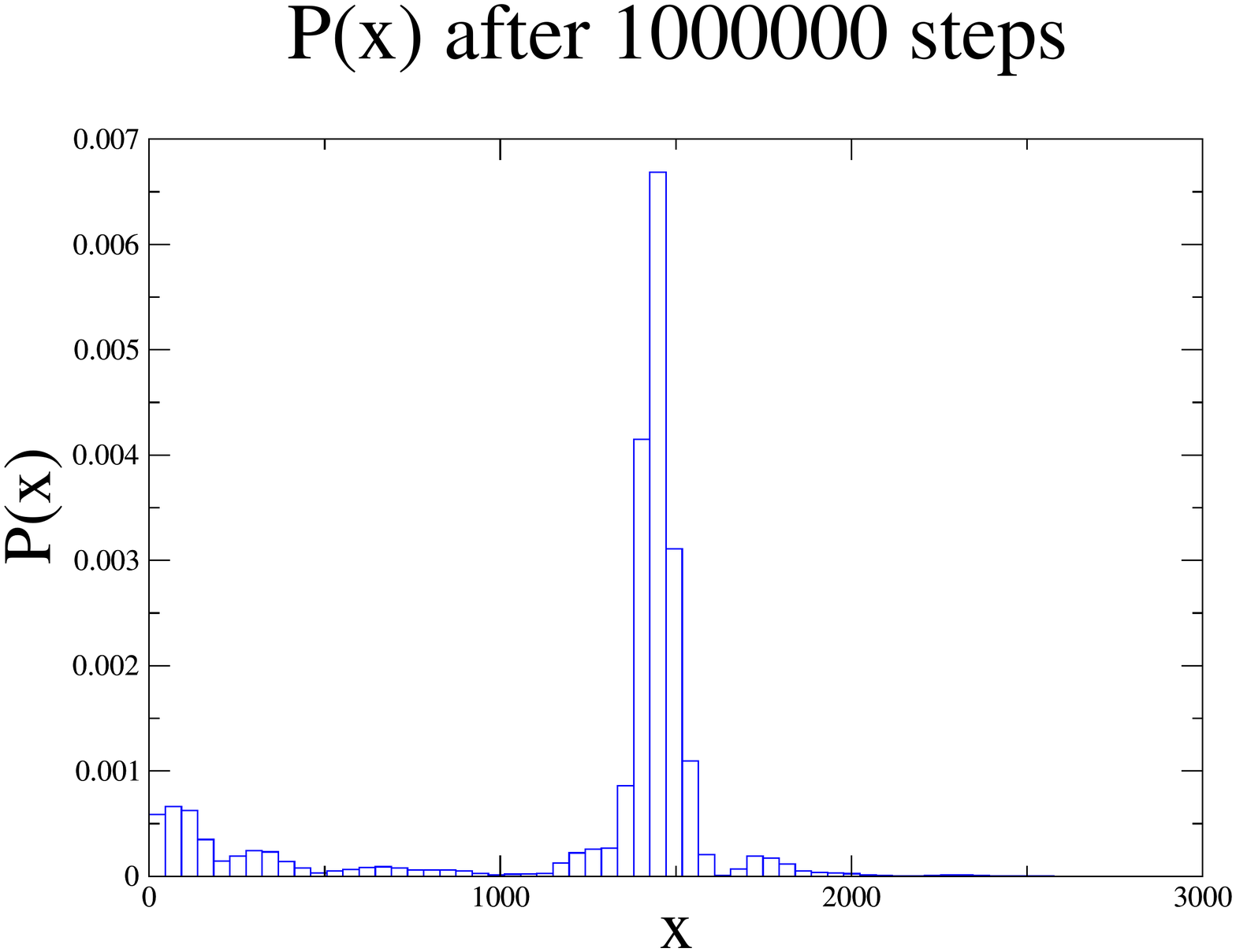}\\
  \caption{Cumulative distribution functions of the position of particle 1 in Monte-Carlo simulations, for different times. A steady state is only achieved for late times. The bottom figure offers a comparison between nucleosomal PDFs in simulations (after 1000000 timesteps), and in the analytical theory.}\label{superman}
\end{figure}

As a validation, we first consider the case treated in Section 3, where the nucleosomes are point-like and only interact via simple exclusion, through their inability to overtake each other. In Fig.~\ref{superman}, we compare the analytically derived positional PDF for the first particle with that measured in Monte Carlo simulation. The two PDFs are in fairly good agreement, although they slightly differ quantitatively in some of the peak heights. This discrepancy is most likely due to (small) differences between the real DNA potential and the fitted one, which is piecewise linear. Indeed, there is a larger difference between the DNA potential and the fitted one between bases 175
0 and 2000, which is also the region where the simulated and analytical PDFs differ the most.

\subsection{Results for the positional PDFs with finite histone size}

In this Section, we study the same polynucleosome chain of $N=10$ nucleosomes, on a DNA with $L=4600$ base pairs, where, however, the histone size is now set to 146 base pairs. This size corresponds to the wrapping of 1.75 turns typical of 10-nm chromatin fibres adopting beads-on-a-string structures. 

\begin{figure}[h!]
  \centering
  \includegraphics[width=0.45\textwidth]{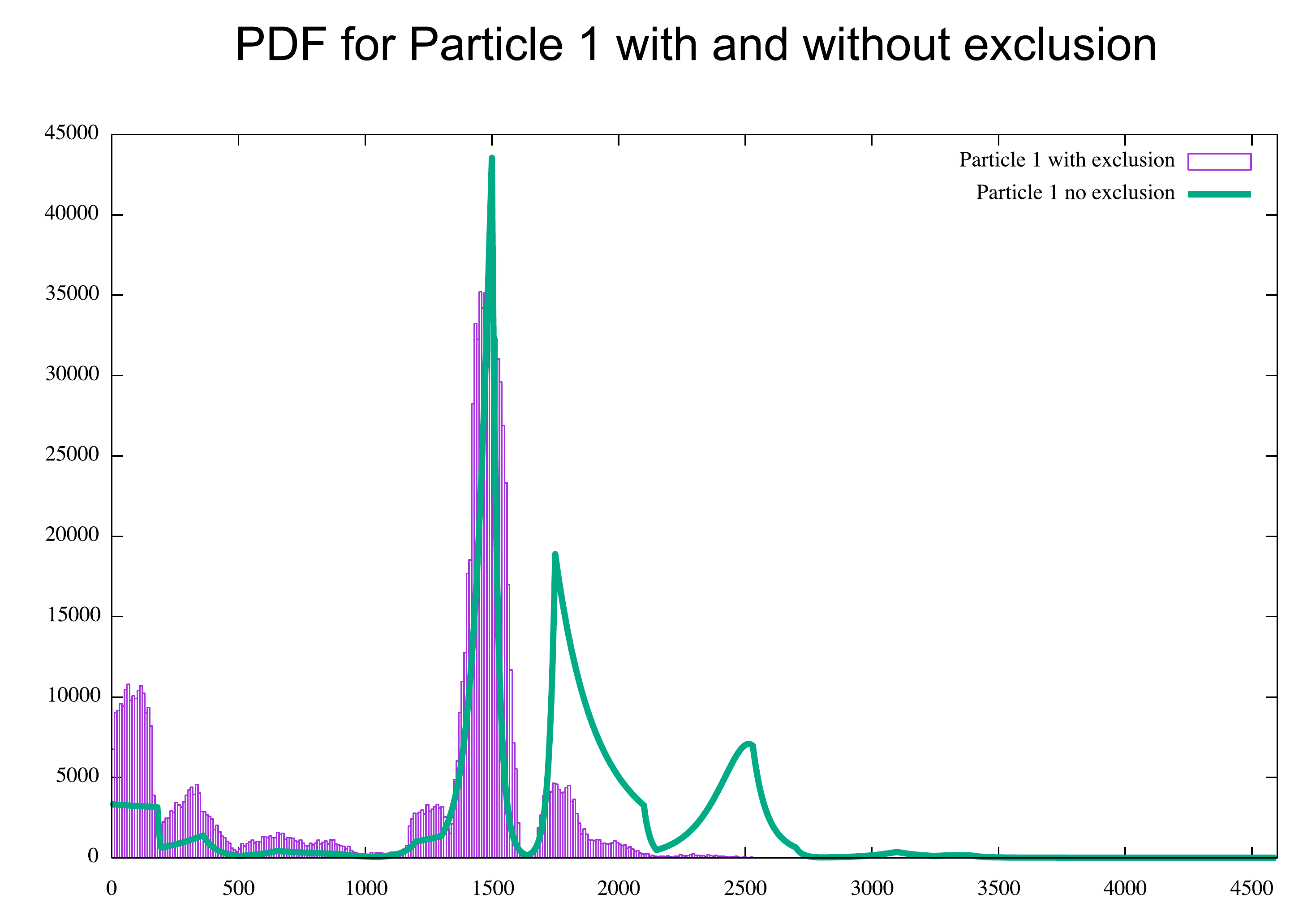}
  \includegraphics[width=0.45\textwidth]{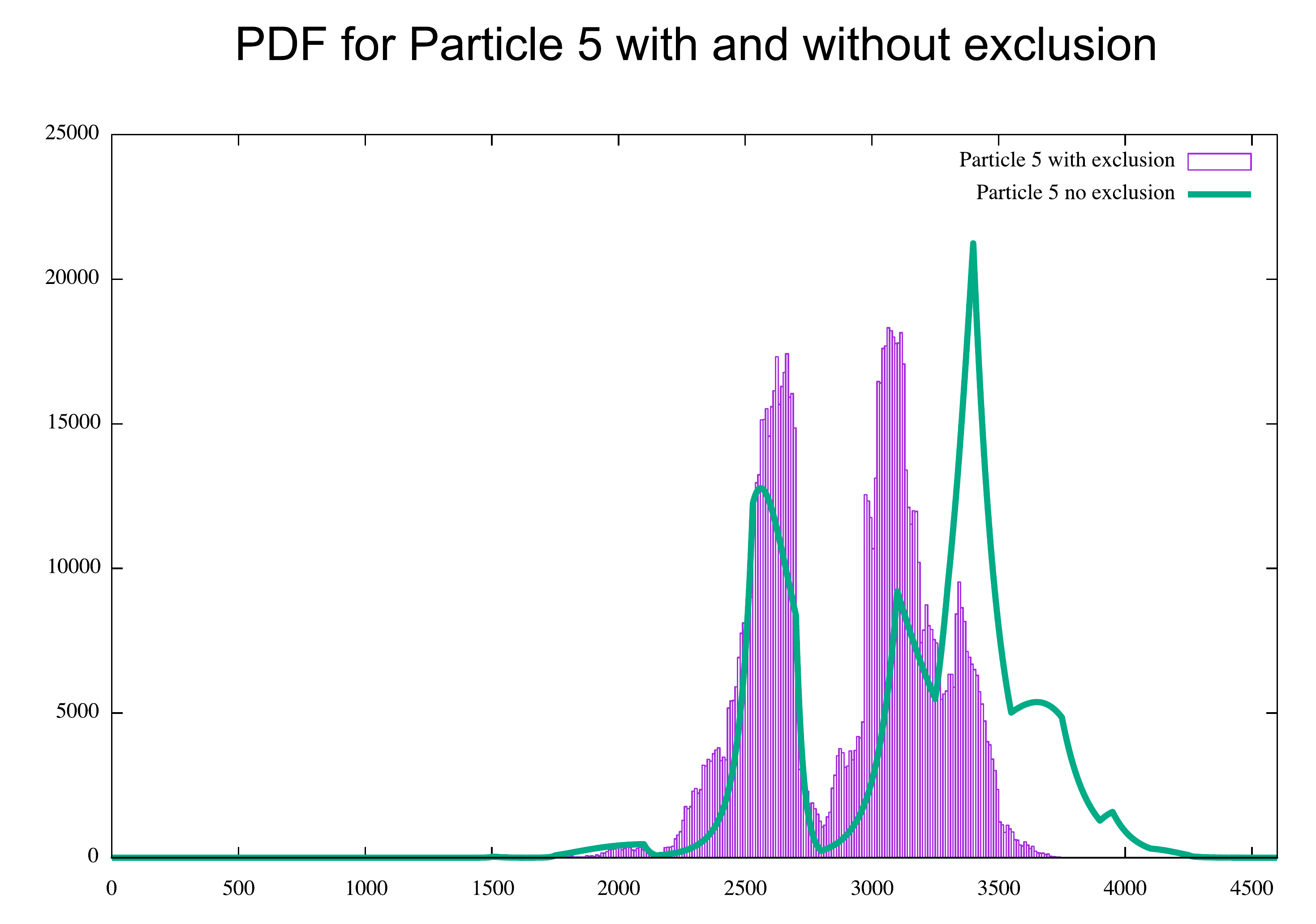}\\
  \includegraphics[width=0.45\textwidth]{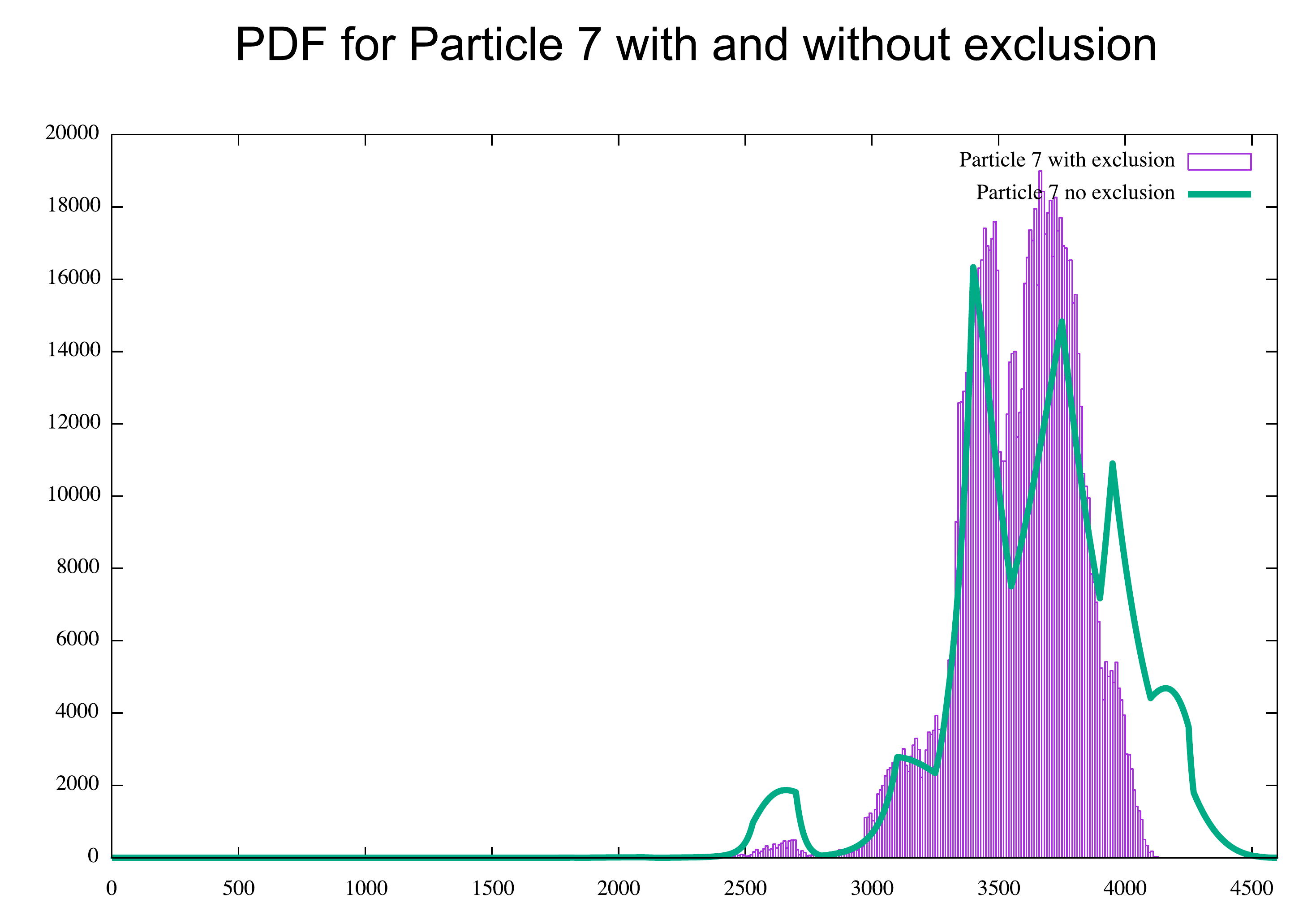}
   \includegraphics[width=0.45\textwidth]{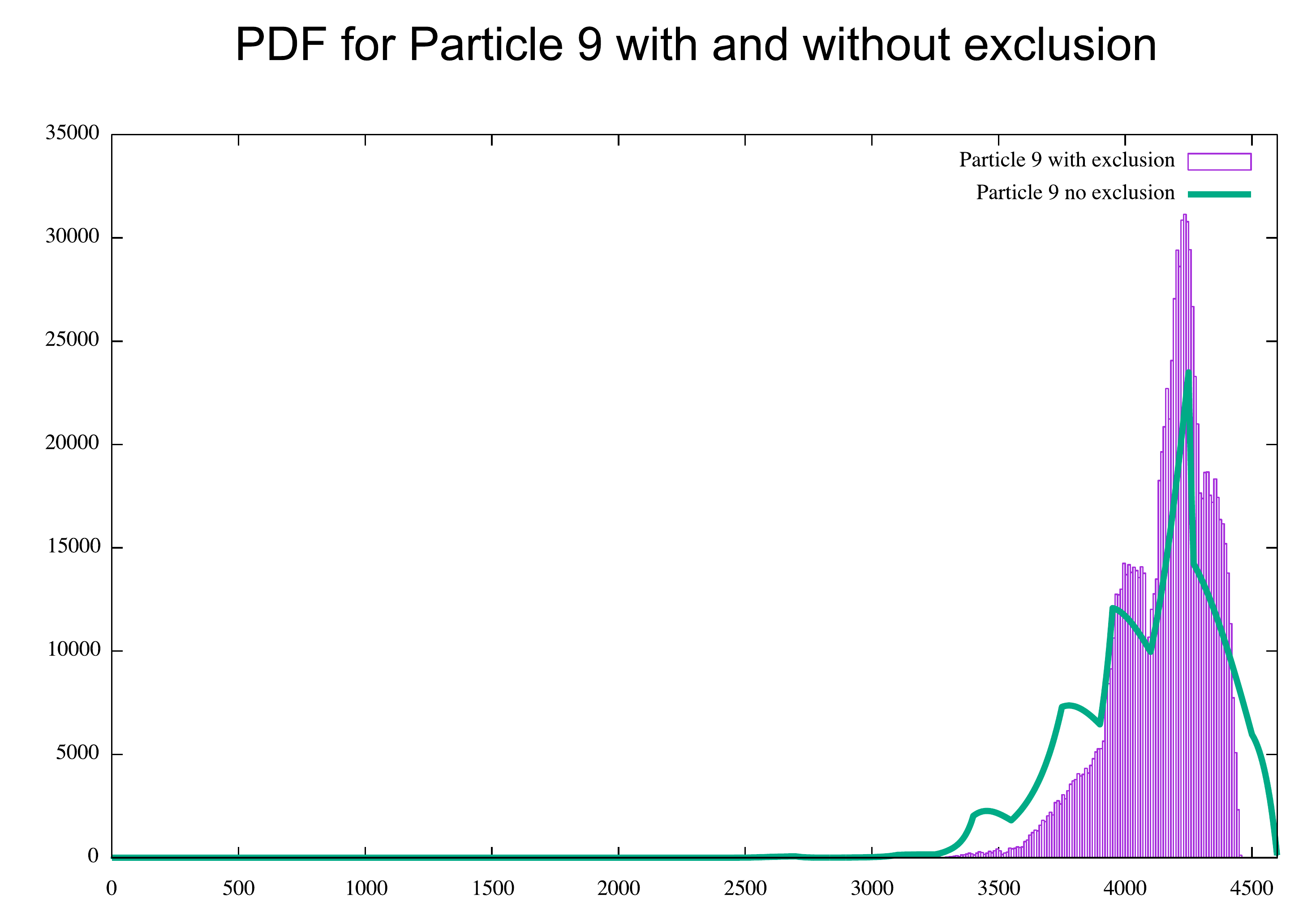}
   \caption{Comparison between analytical PDF for point particles on the genomic DNA potential in green and numerical simulations for particle positionings with excluded volume in purple. The effect of exclusion does not qualitatively change the particles PDF in terms of number of peaks and relative heights but marginally shrinks them as a consequence of the enhanced steric effect.}\label{superman3}
\end{figure}

First, we consider the case of the beta lactoglobulin gene potential. Fig. \ref{superman3} shows the comparison between the analytical particles PDFs and the finite histone size simulations. Pleasingly, there is a good semiquantitative agreement between simulations and theory: this confirms the expectation that our analytical theory works well at low nucleosome density. The density corresponding to $N=10$ is two- to three-fold smaller than the one relevant {\it in vivo} (which is 1 nucleosome/200 base pairs); however, this packing is well in the range explored {\it in vitro} for reconstituted chromatin.

\begin{figure}[h!]
  \centering
  \includegraphics[width=0.45\textwidth]{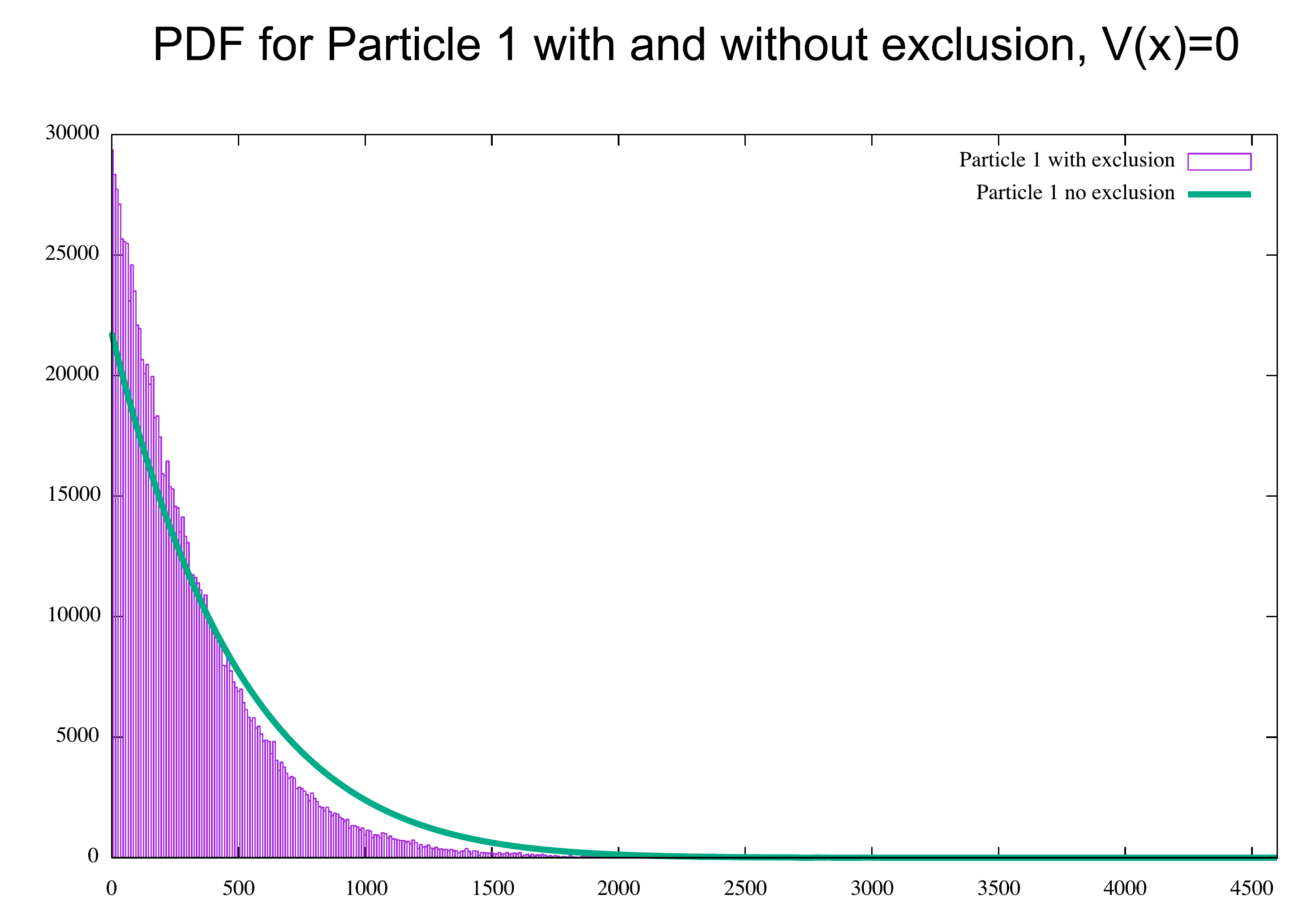}
  \includegraphics[width=0.45\textwidth]{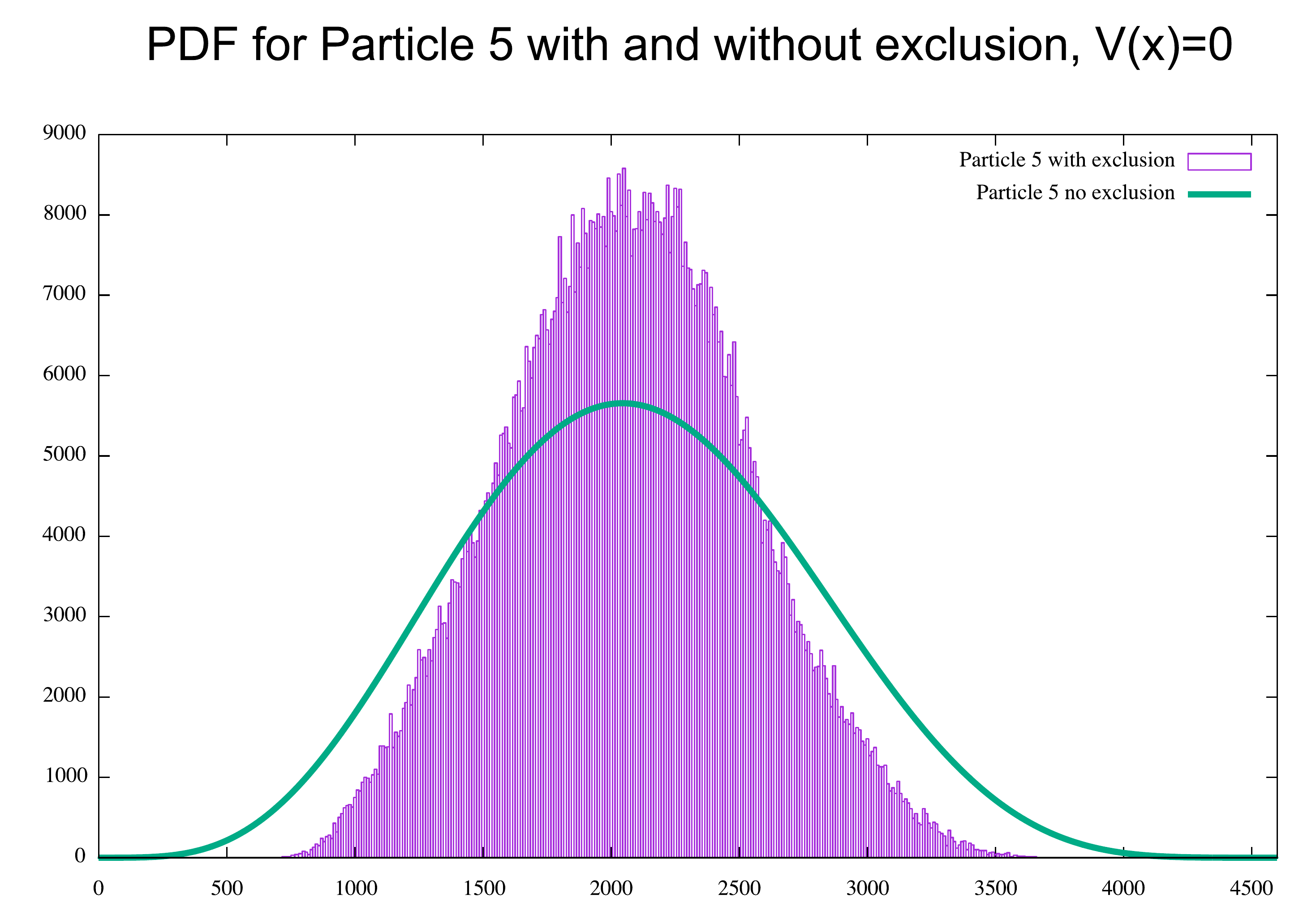}\\
  \includegraphics[width=0.45\textwidth]{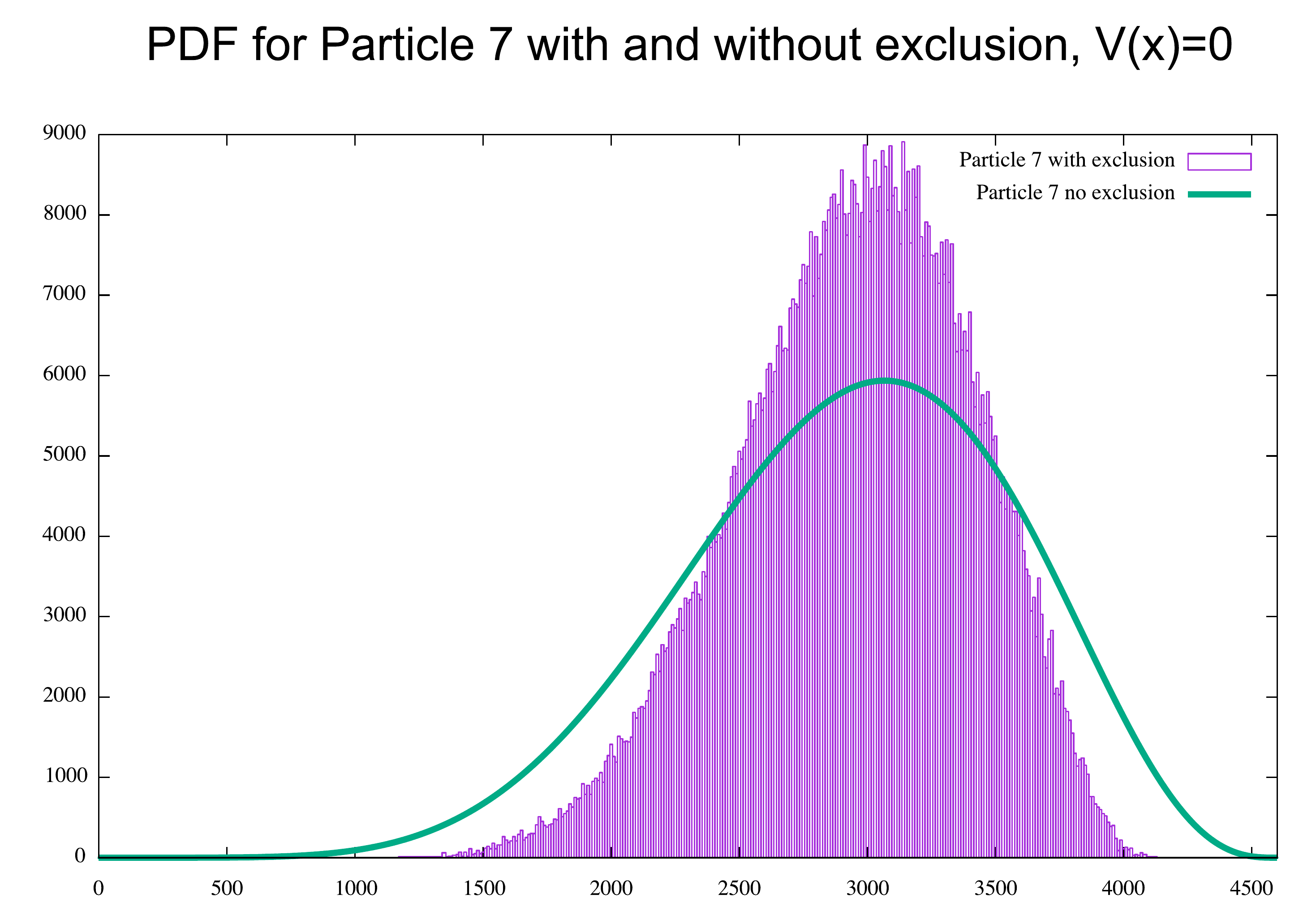}
   \includegraphics[width=0.45\textwidth]{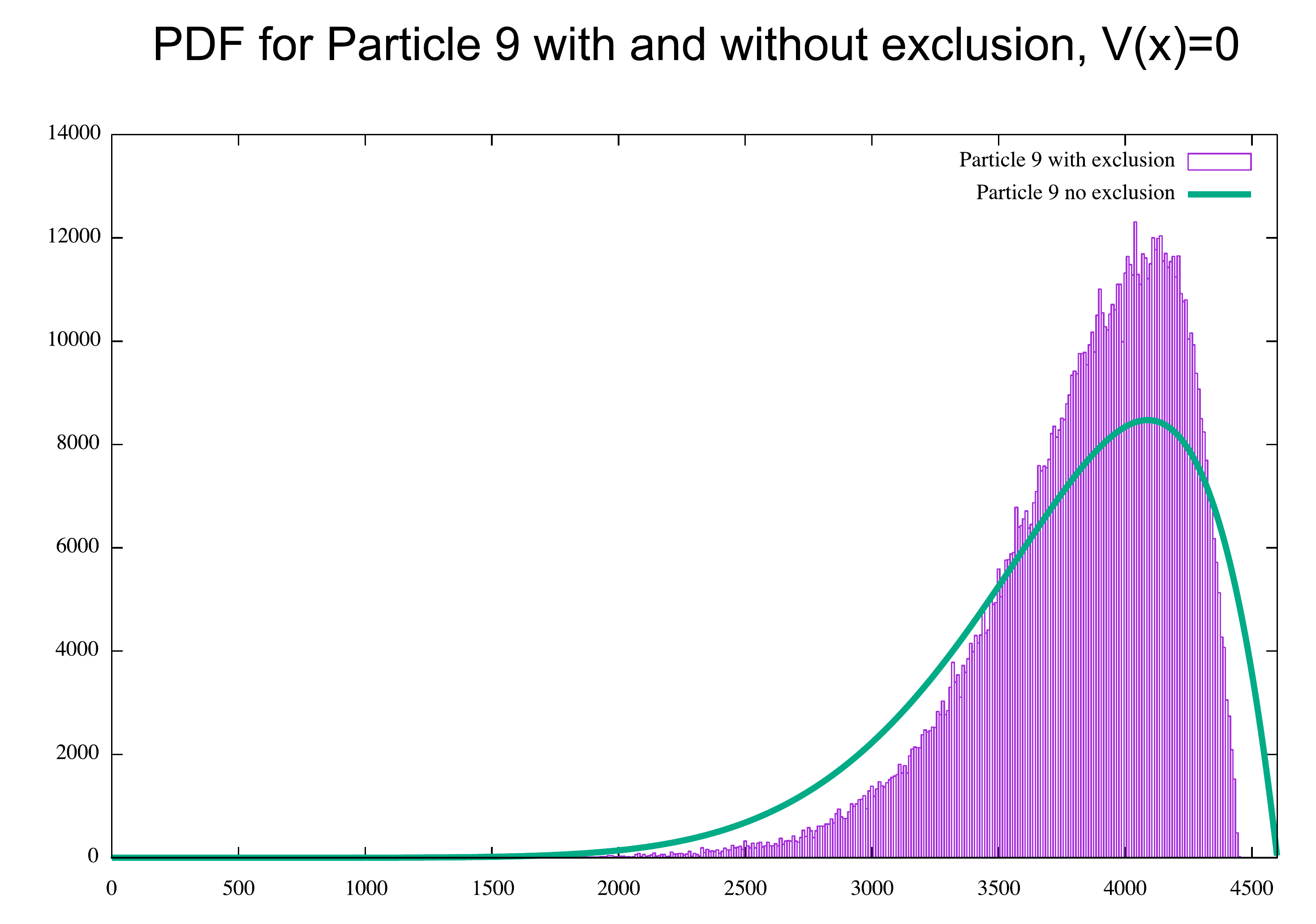}
   \caption{Comparison between analytical PDF for point particles on the homogeneous DNA lattice (green) and numerical simulations accounting for excluded volume (purple).}\label{homogeneousPDF}
\end{figure}

{Similar results are obtained with the homogeneous DNA, where nucleosomes interact only via excluded volume (Fig.~\ref{homogeneousPDF} shows the comparison between simulations and theory for the positional PDF of particle 1).}


\subsection{Simulated digestion patterns}

\begin{figure}[h!]
  \centering
  \includegraphics[width=0.45\textwidth]{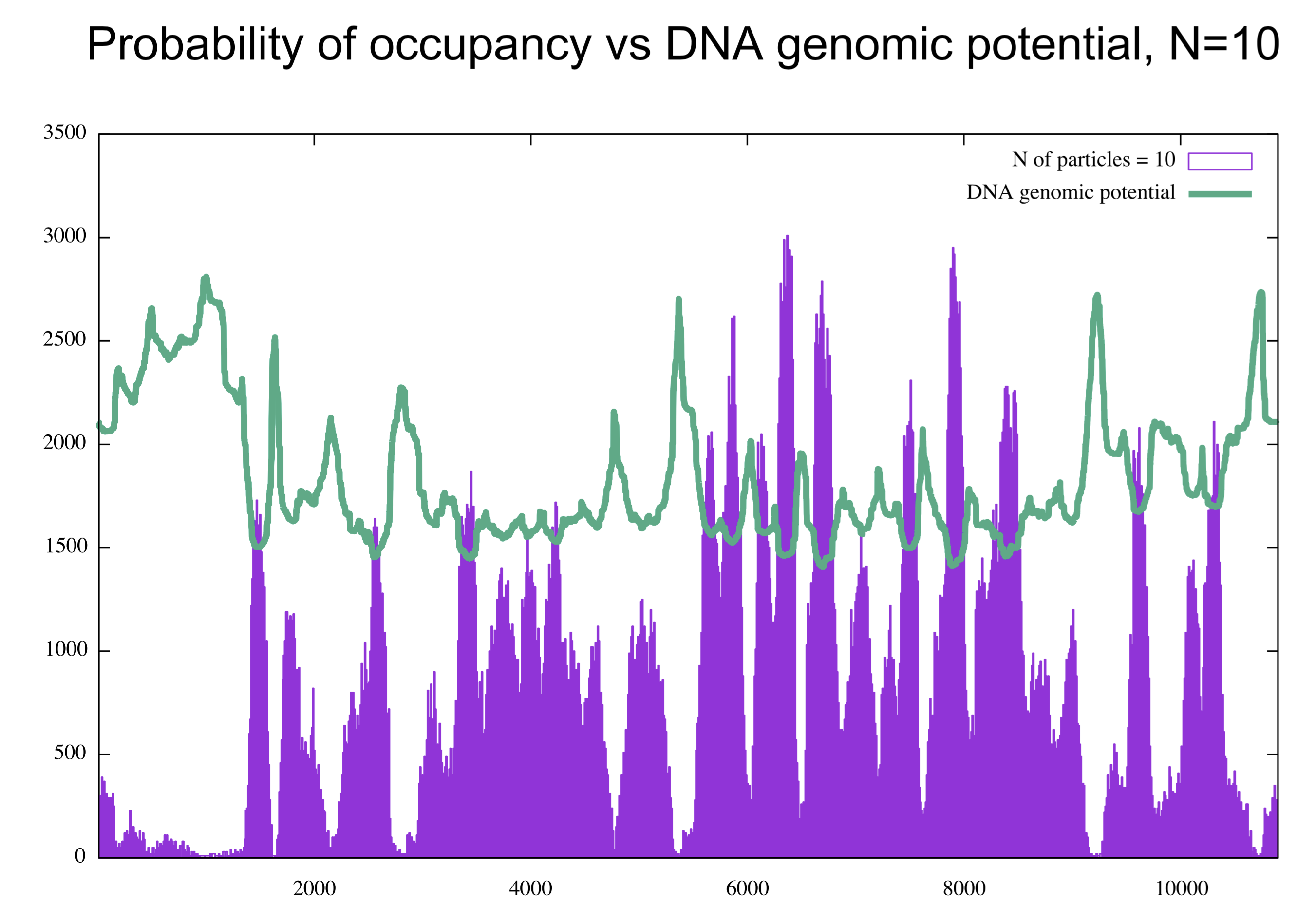}
  \includegraphics[width=0.45\textwidth]{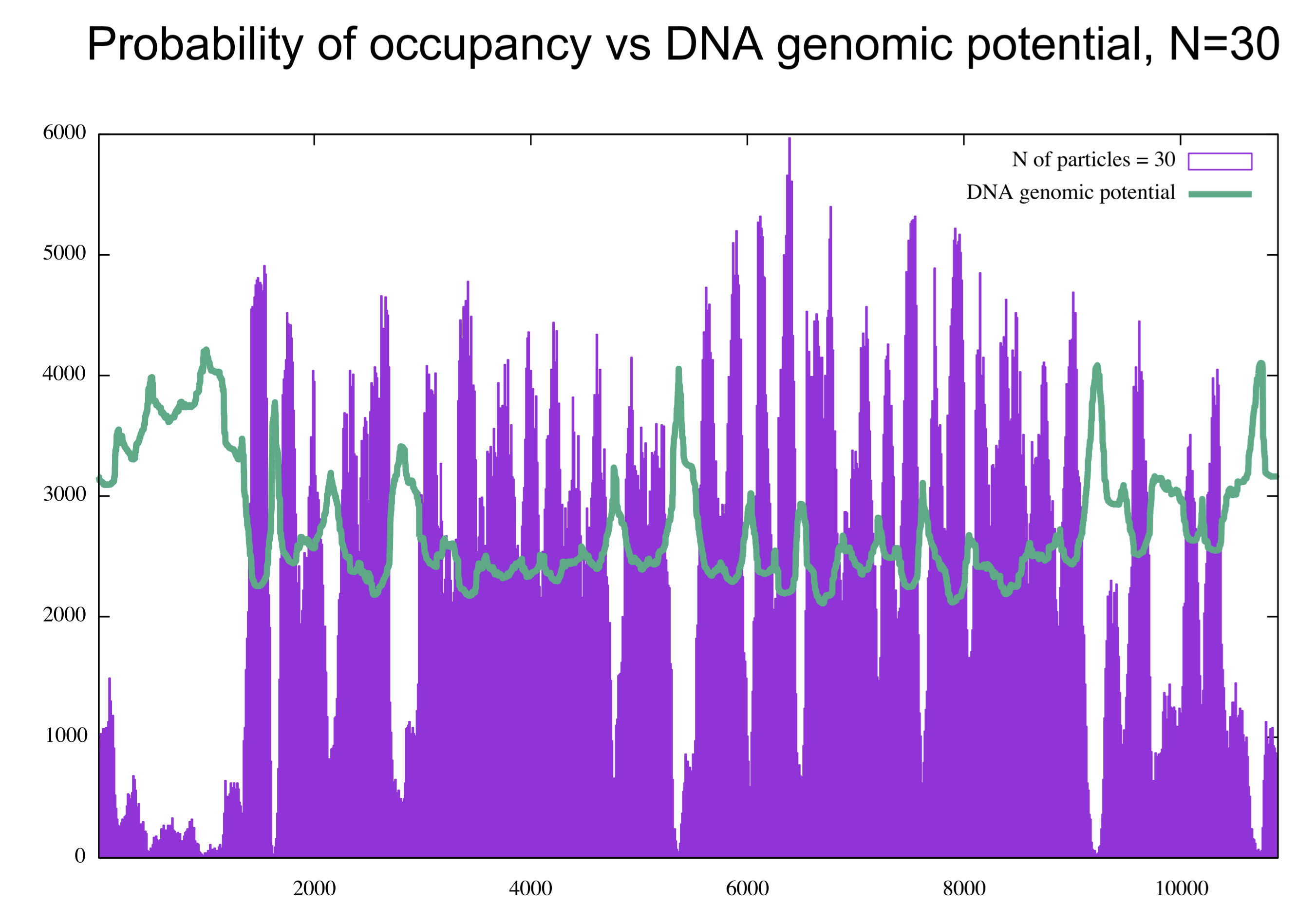}\\
  \includegraphics[width=0.45\textwidth]{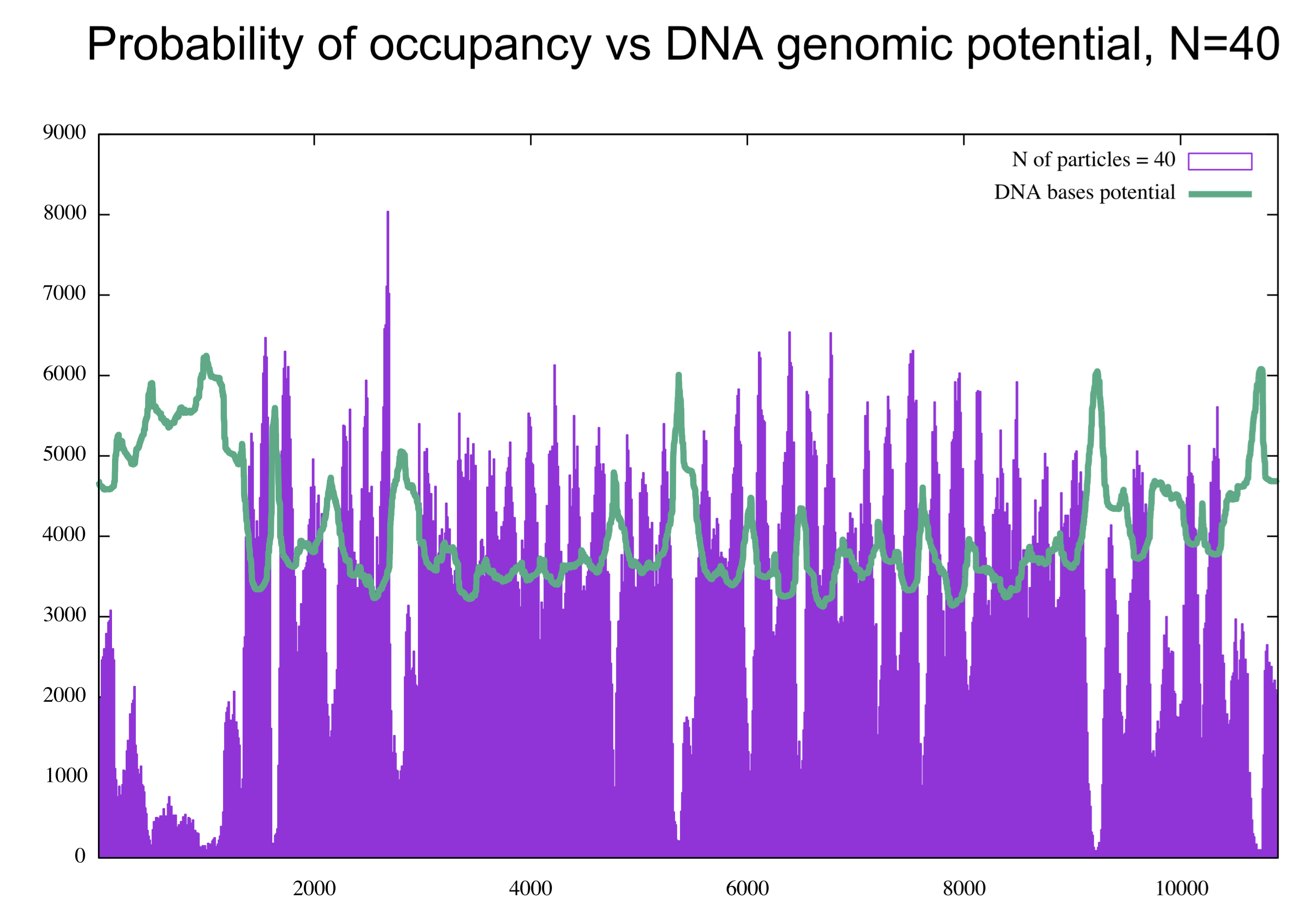}
   \includegraphics[width=0.45\textwidth]{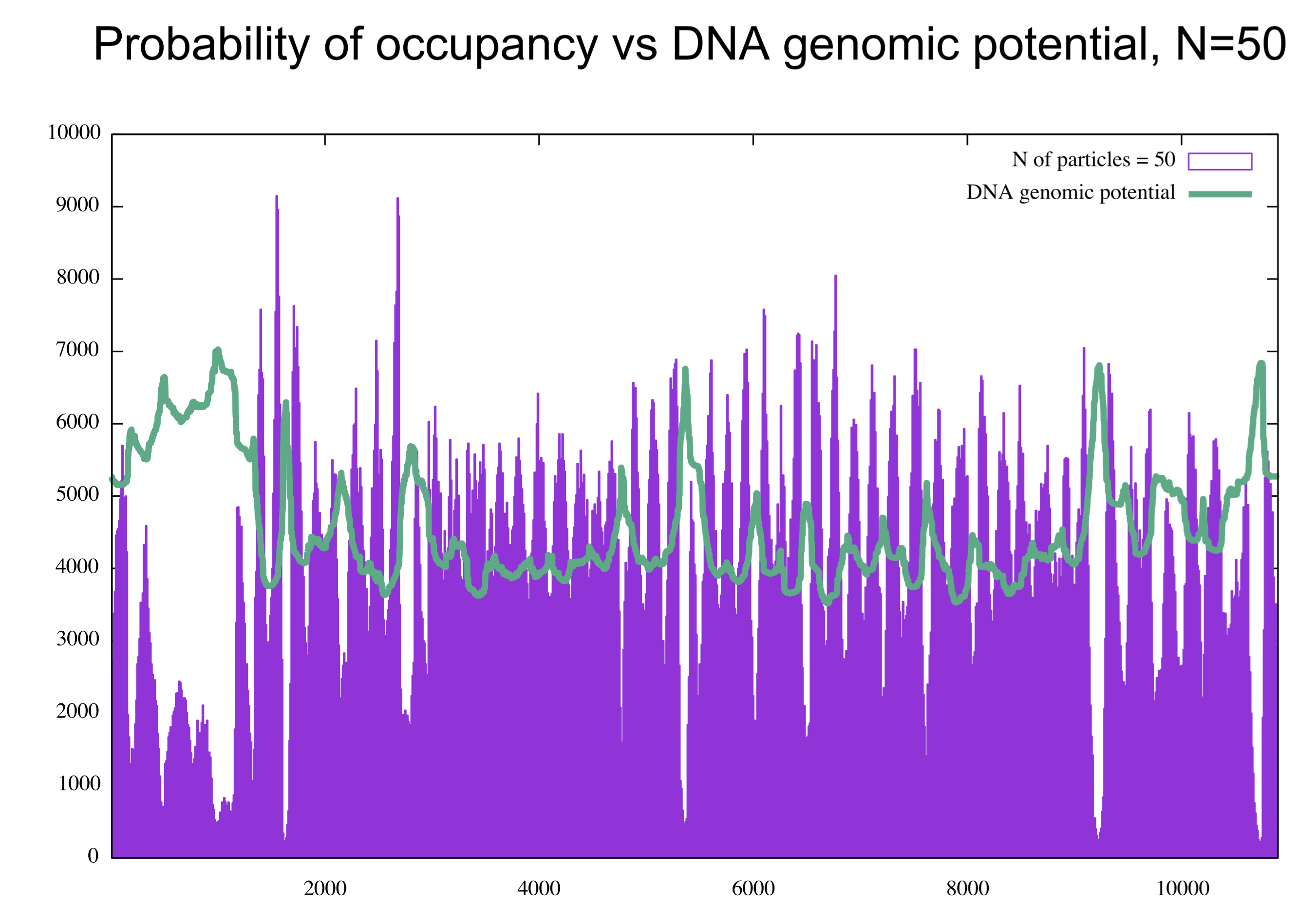}
   \caption{Numerical simulations with periodic boundary conditions yield the probability of histone occupancy with 10, 30, 40 and 50 particle on the DNA lattice. The genomic potential is superposed in green and peaks in occupation probability correspond to troughs in genomic potential. The distributions become sharper for denser crowding.}\label{insilicopositions}     
\end{figure}

Finally, in this Section, we consider polynucleosome loops, with different number of nucleosomes, $N$; we also study the full lactoglobulin gene. In this way we recreate conditions which are closer to experiments on chromatin reconstituted by salt dialysis. Because the DNA is a loop, the positional PDFs of nucleosomes in the homogeneous potential are uniform (i.e., flat) due to translational invariance. The positional PDFs corresponding to the sheet potential are instead shown in Fig.~\ref{insilicopositions} for the various cases considered. Similar considerations qualitatively apply as for the analytical results; especially at low density the nucleosomes are confined close to the potential troughs.

In experiments, it is common to assess chromatin structure and relative nucleosome spacing through nuclease digestion assays~\cite{understandingDNA,invivo,Kornberg1988,chris}.{ \color{black}These experiments consist of two main steps. First, a chromatin fibre is subjected to the action of an enzyme (typically micrococcal nuclease) which cuts `unprotected' DNA: i.e. DNA which is not wrapped up in nucleosomes. This step, known as ``digestion'', leads to a population of DNA fragments of different size: most of these are wrapped around histone octamers, as nuclease quickly degrades naked DNA. Following digestion, the salt concentration is tuned so that histone octamers subsequently unbind. The second main step is then to perform a gel electrophoresis experiment on the remaining fragments of DNA: as the mobility depends on charge, hence length, these experiments give a measure of the size distribution of fragments associated with histone octamers following digestion. The distribution gives a measure of the 1D organisation of the nucleosomes along the fibre: for instance mononucleosomes, dinucleosomes or more complex structures contribute differently to the size distribution as the fragment length associated with those is normally different~\cite{understandingDNA}. It is also of interest to ask whether and to what extent the sequence affects the probability distribution which is measured by these experiments, and that is the question we address here.}

Although the efficiency of micrococcal nuclease depends on local DNA sequence, as a first approximation it is common to model such digestion experiments by assuming that each DNA base pair not associated with a nucleosome is cut with a probability $p$, which depends on the digestion time and efficiency~\cite{Kornberg1988,chris}. We started from this simplified view, and simulated digestion experiments for our polynucleosome chains generated {\it in silico}. In the simulated digestion, we further assumed that only fragments containing at least one nuclesomes remain in the gel (as nuclease quickly degrades unprotected DNA which it has got hold of).

\begin{figure}[h!]
  \centering
  \includegraphics[width=0.95\textwidth]{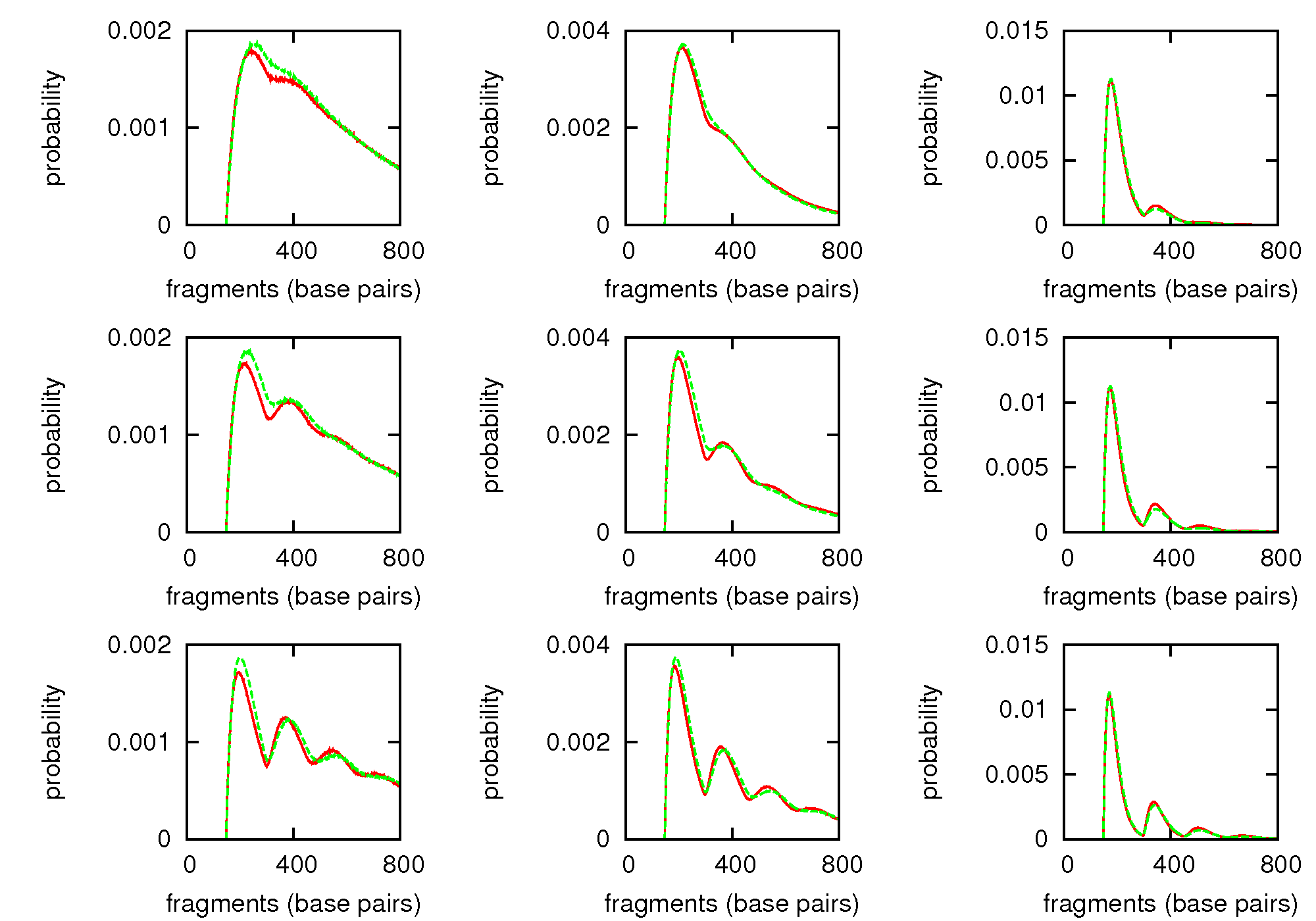}
   \caption{Simulated digestion patterns for polynucleosome chains, for the beta lactoglobulin gene (red), and for a homogeneous DNA (green). Plots correspond to: $N=30$ nucleosomes (left column), $N=40$ (middle column), and $N=50$ (right column). The probability that the nuclease cuts an unprotected base pair is set to $p=0.005$ (top row), $p=0.01$ (middle row) and $p=0.03$ (bottom row). }\label{insilicodigestion}
\end{figure}

Fig.~\ref{insilicodigestion} shows the distributions of fragments resulting from a simulated digestion experiments for a variety of parameters. These plots address the question whether it is possible to detect sequence-dependent nuclesome positioning effects via digestion experiments. Our results show that the sequence signature is extremely sensitive to efficiency, or duration, of digestion, and to nucleosome density. At the same time, quite surprisingly, there is only a very subtle difference between the beta lactoglobulin loops and the homogenous DNA. The largest difference between fragments remaining after digestion with a sequence heterogeneous or homogeneous DNA can be found, according to our simulations, for intermediate densities, and for low or intermediate digestion efficiency. 

{\color{black}These results are non-trivial, but can be rationalised via the following arguments.} At small density, the nucleosomes are so far that nuclease cuts usually contain a single nucleosome, so that sequence effects on gap PDFs are not important. At very high density, the spacing is mainly dictated by steric interactions, hence sequence is again irrelevant. At intermediate densities, instead, the positions of potential minima leads to nontrivial correlations in gap PDFs, and this leads to different digestion patterns. {\color{black}This suggests that sequence effects should be more visible at intermediate density. Furthermore, extreme digestion in our model only leaves nucleosomal DNA, with 146 base pair fragments: the resulting distribution is close to a Dirac delta function, and again sequence effects are washed away. Therefore, low or intermediate digestion and intermediate density are more likely to leave detectable sequence signatures. This is in line with our numerical results, although the quantitative magnitude of the effect (which is difficult to estimate {\it a priori}) is found numerically to be extremely small (see Fig.~\ref{insilicodigestion}).}

\section{Conclusions}

In this work, we have studied a 1D statistical mechanics model of a 1D polynucleosome chain. Our main focus was on the effect of sequence on{ \color{black}the 1D organisation of nucleosomes along the chromatin fibre}, and to address this we have compared the statistics of a DNA molecule with uniform DNA:histone interaction to another case, of a genomic DNA region, where the DNA:histone potential was informed by existing experimental high-resolution nucleosome positioning data. The results we obtain are therefore the interplay between the sequence-dependent potential and excluded volume interactions between nucleosomes within the chain. 

We have first shown that, if we approximate the sequence-dependent potential via a piecewise linear function, we can obtain analytically explicit formulas for the probability distribution functions (PDFs) of each of the nucleosomes within the chromatin fibre. However, our analytical approximation did not consider the effect of the finite size of nucleosomes (which cannot occupy less than 146 base pairs in reality), as, for simplicity, it dealt with pointlike nucleosomes.

We have then presented Monte-Carlo simulations of the model, where finite size effects can be fully taken into account. After validating our algorithm by reproducing the particle PDFs with pointlike nucleosomes, we have studied how the results change when accounting for the finite size of nucleosomes. The general trend is that, as expected, as the nucleosomal density along the fibre increases, the effect of finite size becomes more important; at the same time, for large densities sequence-dependent features become less visible. In practice, it would not be trivial to directly compare predicted PDFs for a polynucleosome chain to experimental data; we therefore simulated a digestion experiment on our polynucleosome chain, as this is a popular assay used to characterise the structure and nucleosome distribution of chromatin fibres in the test tube. The {\it in silico} digestion patterns suggest that there is a possibility to pick up directly sequence effects with such assays, however this requires a careful tuning of the duration of digestion, and of the density of nucleosomes. We hope that these results will stimulate further experimental high-resolution work on nucleosome positioning within DNA of different sequence. It would also be of interest to couple our 1D treatment to simulations of the structure of 3D chromosomes, which are normally done on homogeneous fibres~\cite{chris,chris2,chris3,beard,grigoryev,Arya2009,korolev,perisic}.
\section{Acknowledgments}

We acknowledge J. Allan for useful discussions and for providing us with the nucleosome positioning data on the beta lactoglobulin gene. {\color{black} A. M. acknowledges support form the UK Engineering and Physical Sciences Research Council  (EP/I004262/1)}.  S. T. acknowledges support form the UK Engineering and Physical Sciences Research Council (EP/L504920/1).
\section{Bibliography}

\bibliographystyle{unsrt}                      

\bibliography{PRreferences.bib} 

\end{document}